\documentclass[useAMS,usenatbib]{mn2e}
\usepackage{amssymb}
\usepackage{longtable}
\usepackage{hyperref}
\usepackage{graphicx}
\usepackage{subfigure}
\usepackage{color}
\usepackage{ulem}
\usepackage{amsmath}
\usepackage{stmaryrd}

\title[Centre offsets v.s. velocity dispersion]
   {Correlation between centre offsets and gas velocity dispersion of galaxy clusters in cosmological simulations}
\author[Ming-Hua Li et al.]
   {Ming-Hua~Li,$^1$\thanks{E-mail: liminghua@mail.sysu.edu.cn (LMH);  zhuwshan5@mail.sysu.edu.cn (WSZ)},
  Weishan~Zhu,$^1$\footnotemark[1]
  and Dong~Zhao$^2$
  \newauthor \\
  $^1$School of Physics and Astronomy, Sun Yat-Sen University, 135 Xingang Xi
      Road, Guangzhou 510275, China\\
  $^2$Institute of High Energy Physics, Chinese Academy of Sciences, 19 Yuquan Road, Beijing 100049, China
  }
\date{Draft version}

\pagerange{\pageref{firstpage}--\pageref{lastpage}} \pubyear{2017}

\def\LaTeX{L\kern-.36em\raise.3ex\hbox{a}\kern-.15em
    T\kern-.1667em\lower.7ex\hbox{E}\kern-.125emX}

\def\bc{\begin{centre}}
\def\ec{\end{centre}}
\def\be{\begin{eqnarray}}
\def\ee{\end{eqnarray}}

\begin{document}

\label{firstpage}

\maketitle

\begin{abstract} 
{The gas is the dominant component of baryonic matter in most galaxy groups and clusters. The spatial offsets of gas centre from the halo centre could be an indicator of the dynamical state of cluster. Knowledge of such offsets is important for estimate the uncertainties when using clusters as cosmological probes. In this paper, we study the centre offsets $r_{\rm off}$ between the gas and that of all the matter within halo systems in $\Lambda$CDM cosmological hydrodynamic simulations. We focus on two kinds of centre offsets: one is the three-dimensional PB offsets between the gravitational potential minimum of the entire halo and the barycentre of the ICM, and the other is the two-dimensional PX offsets between the potential minimum of the halo and the iterative centroid of the projected synthetic X-ray emission of the halo. Halos at higher redshifts tend to have larger values of rescaled offsets $r_{\rm off}/r_{200}$ and larger gas velocity dispersion $\sigma_{v}^{\rm gas}/\sigma_{200}$. For both types of offsets, we find that the correlation between the rescaled centre offsets $r_{\rm off}/r_{200}$ and the rescaled 3D gas velocity dispersion, $\sigma_v^{\rm gas}/\sigma_{200}$ can be approximately described by a quadratic function as $r_{{\rm off}}/r_{200} \propto (\sigma_{v}^{\rm gas}/\sigma_{200} - k_2)^{2}$. A Bayesian analysis with MCMC method is employed to estimate the model parameters. Dependence of the correlation relation on redshifts and the gas mass fraction are also investigated.
}
\end{abstract}

\begin{keywords}
 methods: numerical -- galaxies: halos -- galaxies: structure -- dark matter.
\end{keywords}

\section{Introduction}
According to the current structure formation theory in the $\Lambda$CDM universe, galaxies and clusters of galaxies were formed in the potential wells of dark matter halos after the dark age. These halos stem from the primordial density fluctuation which was once amplified by the cosmic inflation in the early Universe. The validity of this structure formation scenario has been well supported by a number of observations and cosmological simulations in the past decades. In this framework, one would expect that ordinary matter  (baryonic matter) is a good tracer of dark matter if the baryonic physics is negligible. 

However, in addition to gravitational interactions, baryon physics including heating, cooling, viscosity dissipation, star formation and feedback will lead to the offsets between dark and baryonic matter in the nonlinear regime. A typical example is the Bullet Cluster 1E0657-558, which exhibits an $8\sigma$ spatial offsets between the $\Sigma$-map (surface density map, observed via X-ray) and $\kappa$-map (convergence map, observed via gravitational lensing) \citep{Bradac2006, Clowe2006}. Knowledge of the offsets between dark and baryonic matter is important not only for an accurate interpretation of  many observed properties of galaxy clusters but also in the estimation of the uncertainties when using clusters as cosmological probes.

For instance, such spatial offsets can lead to the mis-centring problem in the stacked weak lensing study of galaxy groups and clusters, giving rise to biases in mass calibration and the halo mass-concentration relation \citep{john07a, john07b, mandelbaum2010, rozo2011, viola2015}. To reconstruct the underlying mass density field from the tangential shear profile of the background galaxies, one needs to designate the `centre' of groups and clusters, around which shear maps are stacked and azimuthally averaged to compute the surface density around that region. However, a variety of centre candidates lead to ambiguity in the identification. In practice, the brightest group/cluster galaxy (BCG), X-ray peak and centroids of system members are often adopted as the centre. Nevertheless, these adopted centres usually have offsets from the minimum of potential well of associated dark matter halos. Halo mass estimated from this stacked weak lensing method can potentially be biased by $5-30$ per cent if inaccurate halo centres are adopted and the mis-centring issue is not addressed \citep{george2012}. 

In addition, such spatial offsets is important to study galaxy clusters detected by the Sunyaev-Zel'dovich(SZ) effects. Offsets between the centre of gas distribution, X-ray peaks, and BCG may play important role in recovering the SZ flux of clusters \citep{Sehgal2013}. Difference between the distribution of offsets could indicate different selection effects between optical, x-ray and SZ surveys of clusters \citep{Rossetti2016}.

To understand the offsets between BCGs, X-ray peaks and centroids from the minimum of potential of dark matter halo, many efforts have been devoted in observational works. \citet{Lin2004} showed that about $30$ per cent of the BCGs lie at $r>0.05~r_{200}$ from the X-ray peak in a samples of 93 galaxy clusters and groups. \citet{Shan2010} found that 45 per cent of a sample of 38 galaxy clusters have offsets $> 10$ arcsec after studying the projected offsets between the dominant component of baryonic matter centre (measured by X-ray observations) and the gravitational centre (measured by strong lensing). The physical projected separation of BCGs and halo centres were found to $<10-15$ kpc for the majority of clusters, although offsets as large as $> 50$ kpc were reported as indicators of experiencing violent merge \citep{Sanderson2009, Hudson2010, oguri2010, Mann2012, george2012, zitrin2012, Rossetti2016}. The offsets have been shown in simulations as well. \citet{Berlind2003} found that the average separation between the central galaxy and the halo centre of mass is $\sim 0.1$ $r_{200}$. More recently, \citet{Liao2016} showed that baryons and dark matter can be segregated during the build-up process of the halo in non-radiative hydrodynamical simulations, due to different tidal torques experience. \citet{van2013}, \citet{DF2014}, \citet{cui2016} and the references therein also have demonstrated the offsets of  baryonic matter from dark matter inside halos, and hence the presence of centre offsets.

\citet{john07b} introduced a Gaussian distribution, $P(R_{\rm off})=(R_{\rm off}/\sigma_{\rm off}^2)~{\rm exp}^{-R_{\rm off}^2/2\sigma_{\rm off}^2}$, to characterize the underlying spatial offsets between baryons and dark matter, $R_{\rm off}$, in clusters of galaxies\footnote{$\sigma_{\rm off}=0.42$ h$^{-1}$ Mpc  is the dispersion, the value of which is estimated from mock catalogues.}. A more intricate statistical model of $P(R_{\rm off})$ was later proposed by \citet{viola2015}, who included $p_{\rm off}$ being a probability that the central galaxies are offsets from the centre of their host dark matter halos. In both frameworks, $R_{\rm off}$ and the probability $p_{\rm off}$ were treated as free parameters in fitting the observed stacked excess surface density (ESD) profile to the prediction from the Navarro-Frenk-White (NFW) profile of dark matter halo\citep{NFW1996,NFW1997}, in order to estimate the corresponding halo mass. The above works provided statistical descriptions of $r_{\rm off}$, which have been widely used in the stacked weak lensing analysis of ESD in cluster mass calibration \citep{AM2011, Bie2012, george2012, More2015, sereno2015}. \citet{Saro2015} further introduced a form of double Rayleigh function to describe the centre offsets distributions, with one component indicates the small offsets in relaxed systems, and the other component stand for large offsets in systems which had underwent merges recently. This form of function was found to be capable of describing the offsets between the centres of gravitational potential and Sunyaev-Zel'dovich effect (SZE) for about $50, 000$ clusters in the {\it Magneticum} Pathfinder hydrodynamical simulations \citep{Gupta2017}.

The presence of offsets is likely connected to the disturbed dynamical state of clusters, i.e., unrelaxed-ness, and objects with large offsets were expected to be more disturbed \citep{Clowe2006, skibba2011}. It is interesting to probe whether the centre offsets between baryons and the whole halo can be related to some physical properties, such as the virial mass, velocity dispersion, or the temperature of the gas component, etc. The temperature of gas in galaxy groups and clusters, as well as the velocity dispersion, can in principle be obtained or derived from observations, which will reveal the dynamical state of objects. Moreover, the gas is the dominant baryonic component in many galaxy clusters and groups, so that the offsets of gas centre from the halo centre could be a more representative tracer of the spacial offsets between baryonic and dark matter than the BCGs. On the other hand, studying the possible relation between the offsets and the halo properties may help to understand the distributions of offsets. It also has important consequences for accurate interpretation of cluster observations \citep{van2005, DF2014, cui2016}.

In this paper, we study the centre offsets between the baryons and the entire halo system in cosmological hydrodynamical simulations. 
Two types of offsets are investigated: one is the three-dimension offsets between the gravitational potential minimum of the entire halo and the barycentre of the ICM, and the other is the two-dimension projected offsets between the potential minimum of the halo and the iterative centroid of the synthetic X-ray emission of the halo. These offsets are calculated for halos within the range $10^{12.5}{\rm ~h}^{-1}~\textmd{M}_\odot  < M_{200} < 10^{15.0}{\rm ~h}^{-1}~\textmd{M}_\odot $, divided into five mass bins. The correlation between the spatial offsets and the velocity dispersion of the gas within the virial radius, as well as the dynamical state of the system and their evolution since $z=0.5$ are investigated. The consistency of our results with previous observations and simulation works and the possible impacts of the gas fraction of the system are also examined and discussed.

The rest of the paper is organized as follows. In Section 2, we give the details of the simulations and the halo samples, and the different definitions of centres and offsets that are studied in this paper. The results are presented in Section 3.
 The cumulative distributions of the two types of centre offsets and the velocity dispersion of the gas are presented in this Section 3.1 and 3.2. A Bayesian analysis with the Markov-Chain Monte-Carlo (MCMC) method is applied in Section 3.3 to study the correlation between the spatial offsets and the velocity dispersion of the gas as well as its redshift evolution. Conclusions and discussion about the results are presented in Section 4.

\section{Methodology}
\subsection{Numerical simulations}
The Tree-PM $N$-body/SPH code GADGET-2 \citep{springel2005} is used to carry out the simulations in this work. The simulations adopt a flat $\Lambda$CDM cosmology, giving the concordance parameters as $\Omega_{\rm m,0} = 0.30$, $\Omega_{\rm \Lambda, 0} = 0.70$, $\Omega_{\rm b,0} = 0.045$ for the matter, the dark energy, and the baryonic matter density parameter respectively, $h = 0.70$ for the current dimensionless Hubble parameter, and $n_{\rm s} =0.96$ for the spectrum index of primordial perturbation \citep{Planck2015}. Multiple physical processes are included in the simulations, including an ultra-violate (UV) background, radiative cooling, star formation, and the feedback from supernovae. Feedback from AGN is ignored at the current stage. 

Our study is based on three simulations started from $z_{\rm ini} = 120$. Each simulation contains $512^3$ dark matter and $512^3$ gas particles. The first two simulations track the structure formation in a periodic box of  $(100$ h$^{-1}$ $\rm {Mpc})^3$. A series of previous work was based on these two simulations, i.e. \citet{Dong2014, Wang2014, Tang2018}. For the first run, which is referred as the `B100-SF1' (`SF1' for short) in the rest of the paper, the adopted Plummer equivalent softening length  is $\epsilon_{{\rm Pl}} = 9.8$ h$^{-1}$ kpc . The second simulation, referred as B100-SF2(SF2 for short), inherits most of the parameters of the `SF1' run but adopts a different softening length $\epsilon_{{\rm Pl}} = 4.5$ h$^{-1}$ kpc to examine the possible influence from the resolution.

The third simulation (`B200-SF3', `SF3' for short) adopts most of the input parameters of the `SF2' run, but with a larger box size of $(200$ h$^{-1}$ $\rm {Mpc})^3$, which would increase the number of halos with mass over $10^{14.5}{\rm ~h}^{-1}~\textmd{M}_\odot$. We apply the Amiga's Halo Finder (AHF, \citet{KK2009}) to identify haloes from the simulations as a mean density $\rho(<r_{200}) = 200 \rho_{\rm crit}$ within virial radius, where $\rho_{{\rm crit}}\equiv 3H_0^2/(8\pi G)$ is the critical density of the universe. Only the halos with a number of particles larger than $1000$ are considered in our study. 

\subsection{Definition of centres and offsets}
\label{section2.2}
As pointed out by \citet{robertson2017}, the types and magnitudes of the centre offsets in clusters sensitively depends on different definitions of the halo centres. For the offsets calculation in this work, the following centre definitions are adopted respectively for the entire halo and the baryonic component.

\subsubsection{Halo centre identification}
\label{section2.2.1}
As in many of the literatures, we use the following definition of halo centre.
$\\$

\noindent {\bf Minimum of the Gravitational Potential $\mathbf{r}^{\rm halo}_{{\rm pot}}$:} The minimum of the potential well is used as the most popular indicator of the cluster centre in both the simulation and the weak lensing observation. The potential minimum position of all particles within the virial radius in the comoving coordinate system is denoted as $\mathbf{r}^{\rm halo}_{{\rm pot}}$.

\subsubsection{Baryonic centre identification}
\label{section2.2.2}
When we refer to the baryonic centre in this study, we focus on the ICM gas component. The related definitions of 
the baryonic centres in our investigation of offsets are: 
$\\$

\noindent {\bf Barycentre of the Gas $\mathbf{r}^{\rm gas}_{\rm com}$:} We calculate the centre of mass of all the gas particles, i.e. $\mathbf{r}^{\rm gas}_{\rm com}$. We restrict the calculation to the gas particles that are lying within the halo in the comoving coordinate system.
$\\$

\noindent {\bf X-ray Peak of the ICM $\mathbf{r}^{\rm ICM}_{\rm xpeak}$:} To compare with the X-ray observations of clusters, we also simulate the X-ray emission of the ICM component of each halo. The synthetic X-ray photons are generated from an ideal emission spectrum that is calculated for each gas particle depending on their temperature, metallicity, and redshift. This procedure is implemented by the PHOX code \citep{Biffi2012,Biffi2013}, which invokes the XSPEC package \footnote{https://heasarc.gsfc.nasa.gov/docs/xanadu/xspec/} to deal with the emission (APEC model, \citealt{Smith2001}) and absorption (WABS model, \citealt{MM1983}) of the X-ray photons. These generated X-ray photons are then filtered according to the specified collecting area ($\sim$ 2000 cm$^2$) and a realistic observing time ($\sim 5\times10^4$ seconds) and finally projected in the direction of line of sight (l.o.s) to obtain the synthetic 2D X-ray map of each halo. The X-ray peak $\mathbf{r}^{\rm ICM}_{\rm xpeak}$, which is identified as the pixel with the maximum value of photon counts, which seems is more close to the observation.

$\\$
\noindent {\bf Iterative Centroid/Geometric Centre of the X-ray Photons $\mathbf{r}^{\rm ICM}_{\rm xgeo}$:} We use an iterative method to calculate the centroid of the X-ray profile of the halo which is similar to that adopted by \citet{cui2016}. For each halo, we start from the X-ray peak, $\mathbf{r}^{\rm ICM}_{\rm xpeak}$, calculate the geometric centre of the X-ray photons within the virial radius, and set it as the initial value. Then in each iteration, we search the centre of X-ray, starting from the position of the centroid obtained in the last step, but with the radius $r$ shrinking to be 0.85 times of previous step. The iteration ends until the searching radius reaches $r<0.3 r_{\rm 200}$. This method and centroid definition, which is based on those in \citet{Bohringer2010} and \citet{Rasia2013}, is believed to be less biased by the satellites than the X-ray peak $\mathbf{r}^{\rm ICM}_{\rm xpeak}$ (see \citet{Mantz2015} for more discussions).
We use the above iterative approach to locate the centroid of the X-ray profile of the halo, which is denoted as $\mathbf{r}^{\rm ICM}_{\rm xgeo}$.

\begin{table*}
\begin{tabular}{|c|c|c|c|c|c|c|c|c|c|}
\hline
 & \multicolumn{3}{|c|}{B100-SF1} & \multicolumn{3}{|c|}{B100-SF2} & \multicolumn{3}{|c|}{B200-SF3}\\ 
& $z=0$ & $z=0.2$ & $z=0.5$ & $z=0$ & $z=0.2$ & $z=0.5$ & $z=0$& $z=0.2$ & $z=0.5$\\ 
\hline
{\bf{M1}}: $M_{200} \geq 10^{14.5}{\rm ~h}^{-1}~\textmd{M}_\odot$ & 3 & 1 & 0 &3 & 1 & 0 & 35 & 25 & 21\\
{\bf{M2}}: $10^{14.0}{\rm ~h}^{-1}~\textmd{M}_\odot  \leq M_{200} < 10^{14.5}{\rm ~h}^{-1}~\textmd{M}_\odot $ & 15 & 12 & 8  & 15 & 12 & 8 & 188 & 167 & 112\\
{\bf{M3}}: $10^{13.5}{\rm ~h}^{-1}~\textmd{M}_\odot  \leq M_{200} < 10^{14.0}{\rm ~h}^{-1}~\textmd{M}_\odot $ & 63 & 66 & 52 & 63 & 65 & 55 & 839 & 802 & 702\\
{\bf{M4}}: $10^{13.0}{\rm ~h}^{-1}~\textmd{M}_\odot  \leq M_{200} < 10^{13.5}{\rm ~h}^{-1}~\textmd{M}_\odot $ & 259 & 232 & 217 & 259 & 241 & 220 & 2649 & 2613 & 2521\\
{\bf{M5}}: $10^{12.5}{\rm ~h}^{-1}~\textmd{M}_\odot  \leq M_{200} < 10^{13.0}{\rm ~h}^{-1}~\textmd{M}_\odot $ & 800 & 803 &750 & 807 & 812 & 768 &7892 & 8019 & 8025\\
\hline
\end{tabular}
\caption{Number of halos within different mass bins for the three cosmological simulations.}
\label{tab1}
\end{table*}

\subsubsection{Definitions of offsets}
\label{section2.2.4}
In this work, we focus on two types of offsets $r_{{\rm off}}$.

$\\$
\noindent {\bf Potential Minimum-Barycentre offsets (`PB' offsets):} It is the offsets between the barycentre of the gas and the gravitational potential minimum of the entire cluster, i.e. 
\begin{center}
\begin{equation}
r_{{\rm off, PB}}= |\mathbf{r}^{\rm halo}_{{\rm pot}} - \mathbf{r}^{{\rm gas}}_{{\rm com}}|.
\label{roff1}
\end{equation}
\end{center}
This is an physically intrinsic offsets between the gas and the entire system and likely can be related to the physical properties of groups and clusters. 

$\\$

$\\$
\noindent {\bf Potential Minimum-X-ray Centroid offsets (`PX' offsets):} The second type of offsets that is studied in this work is the offsets between the centroid of the ICM and the gravitational potential minimum of the entire cluster, i.e. 
\begin{center}
\begin{equation}
r_{{\rm off, PX}} = |\mathbf{r}^{\rm halo}_{{\rm pot}} - \mathbf{r}^{{\rm ICM}}_{{\rm xgeo}}|.
\label{roff2}
\end{equation}
\end{center}
Different from the centre-of-mass of the gas, these X-ray emission centres of the ICM gas could be easily obtained from the X-ray observations of clusters and galaxies. Another possible choice of the baryonic centre is the BCGs. However, since the gas component is believed to dominate over the stellar component in galaxy clusters, the offsets of the gas centre from the halo centre could be a more representative tracer to the spacial offsets between baryonic and dark matter than the BCGs. Considering these, we base our study and discussions on the above two types of centre offsets for the halos in the simulations.

\subsubsection{Different mass and redshift bins}
\label{section2.2.5}
To study the dependence of the offsets $r_{{\rm off}}$ on halo mass, the halos identified in the simulations are classified into five bins according to their virial mass $M_{200}$, with a bin width of $\Delta {\rm log}(M_{200}/{\rm ~h}^{-1}~\textmd{M}_\odot ) = 0.5$. 
The number of halos in different mass and redshift bins are given in Table \ref{tab1}.

We perform our statistical analysis for three different snapshots (different redshifts), i.e. at $z= 0, 0.2$ and $0.5$ to examine the possible redshift evolution of the results. The halos at $z= 0.2$ and $0.5$ are identified by using the same spherical overdensity method as those at $z= 0$.

\begin{figure*}
\subfigure[~\textsf{results for PB offsets}] { \label{fig1a}
\scalebox{0.36}[0.36]{\includegraphics{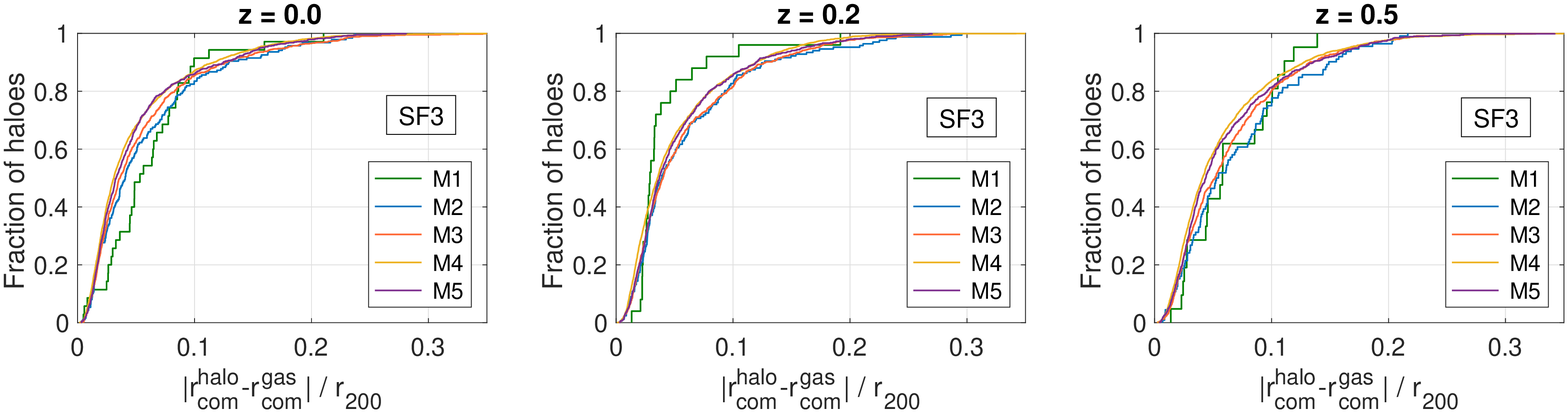}}
}
\subfigure[~\textsf{results for PX offsets}] { \label{fig1b}
\scalebox{0.40}[0.40]{\includegraphics{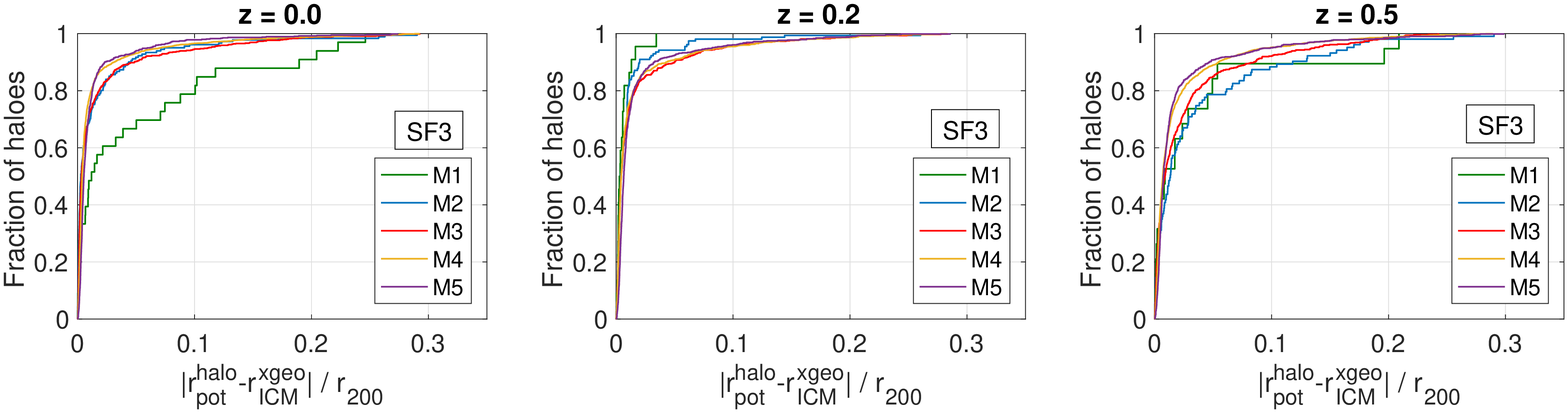}}
}
\caption{Cumulative distributions of the rescaled PB offsets $r_{\rm off, PB}/r_{200}$ and PX offsets $r_{\rm off,PX}/r_{200}$ at three different redshifts for the SF3 run. The offsets distance is rescaled by virial radius $r_{200}$. Lines of different colors refer to different halo mass bins as detailed in Table \ref{tab1}.}
\label{fig1}
\end{figure*}

\begin{figure*}
\includegraphics[width=65mm]{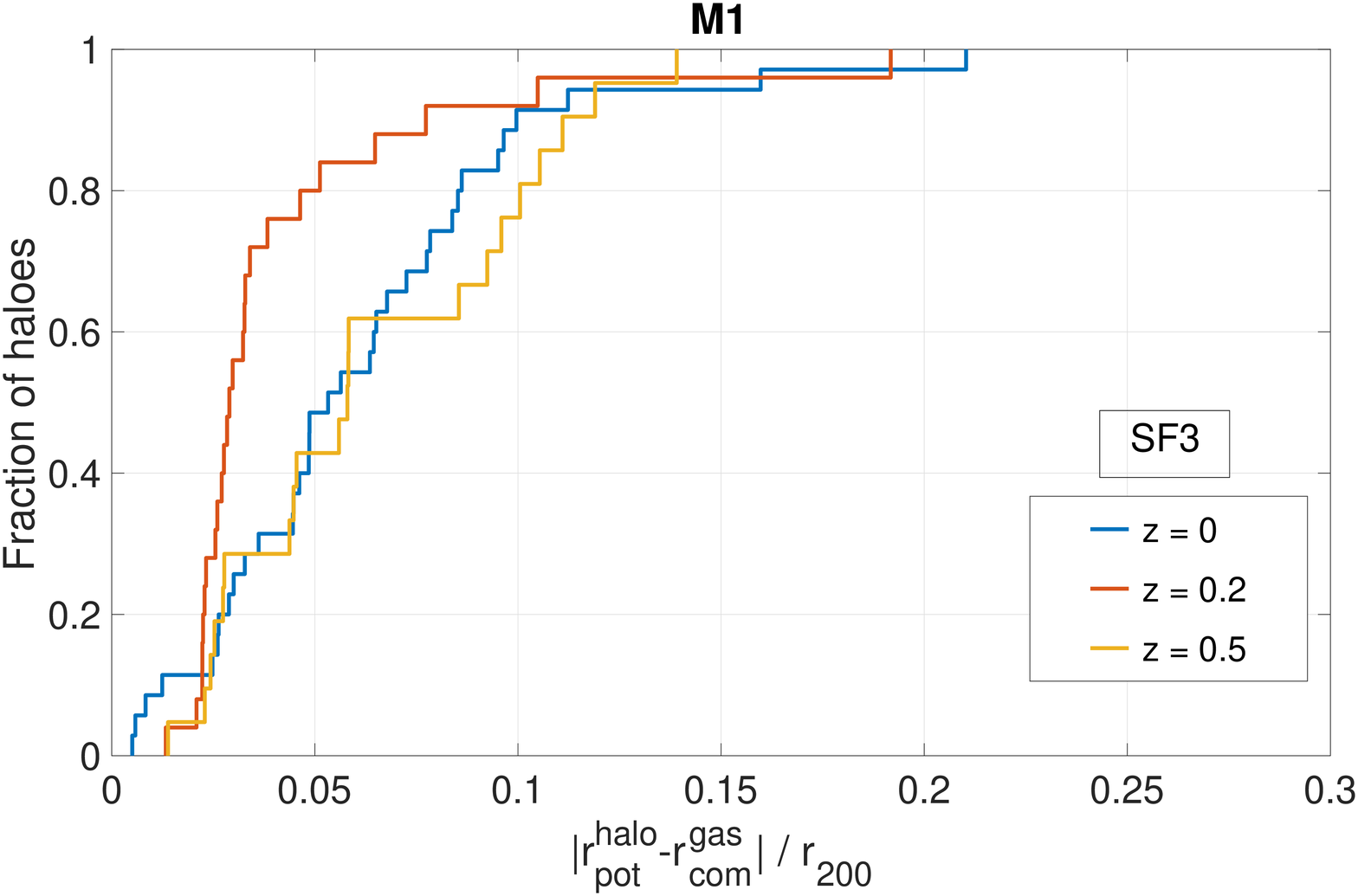}
\includegraphics[width=135mm]{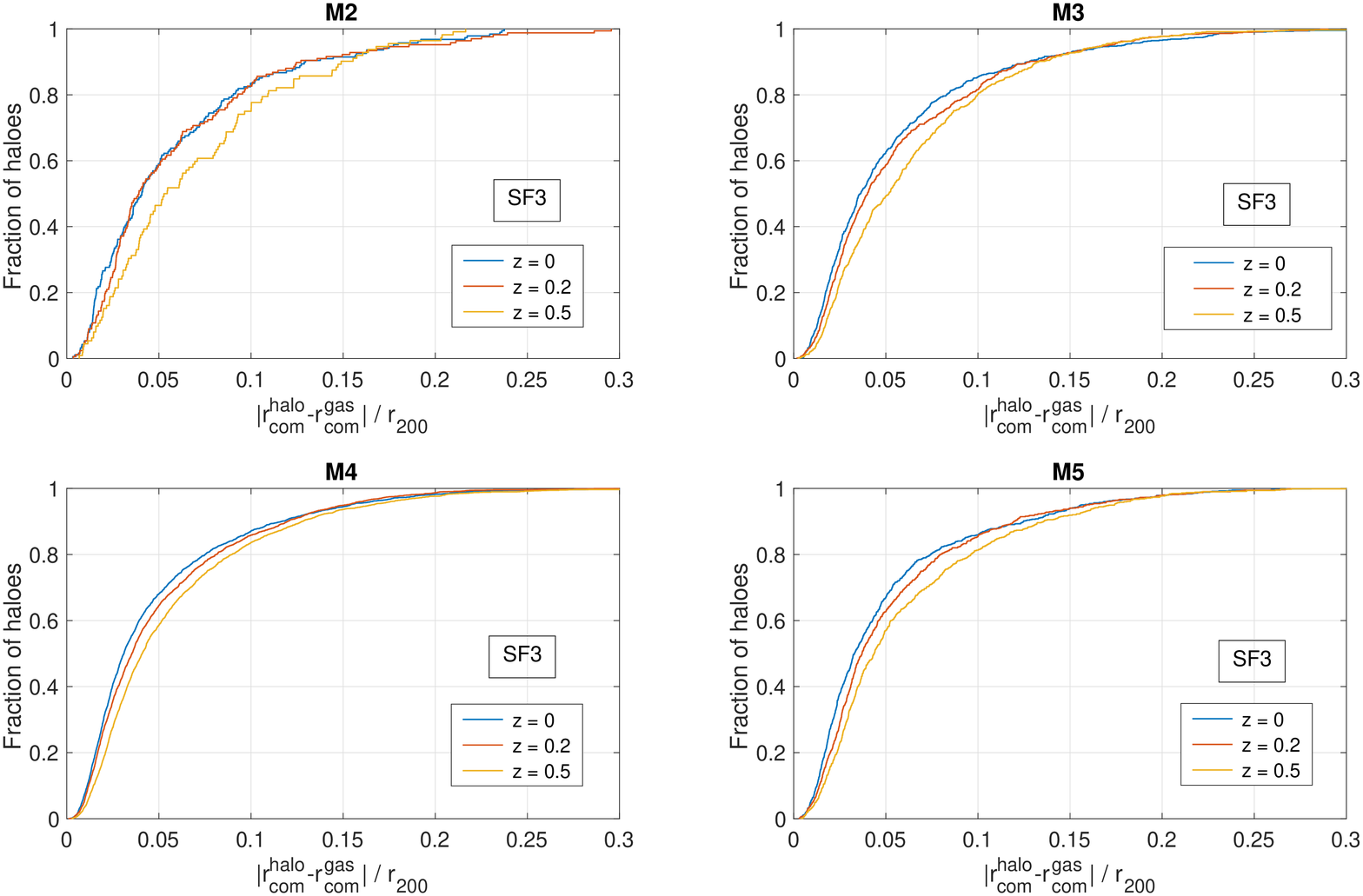}
\caption{Redshift evolution of the cumulative distributions of the rescaled PB offsets $r_{{\rm off,PB}}/r_{200}$ for different mass bins in the SF3 run. Lines of different colors refer to the results for different redshifts.}
\label{fig2}
\end{figure*}

\begin{figure*}
\includegraphics[width=58mm]{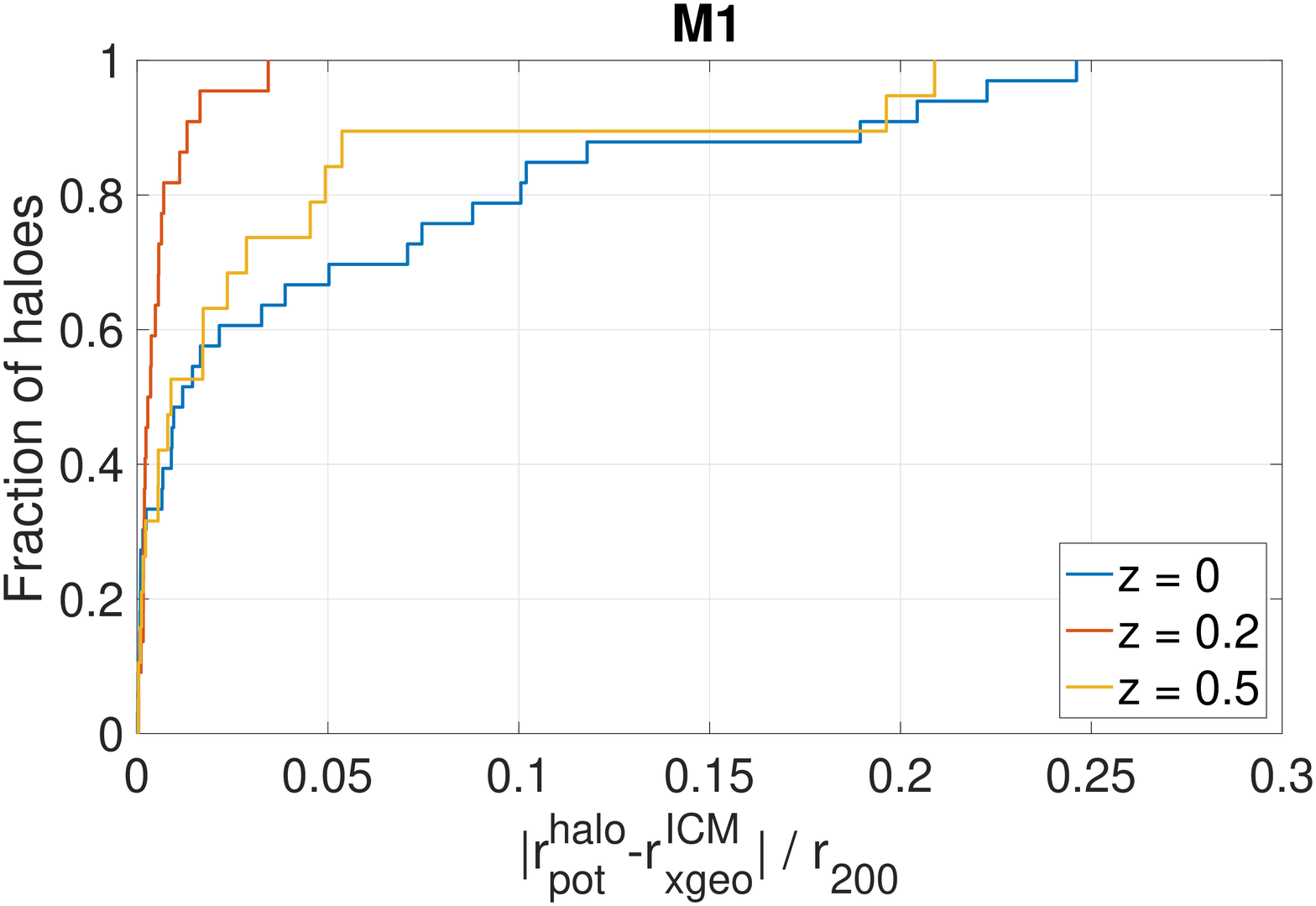}
\includegraphics[width=135mm]{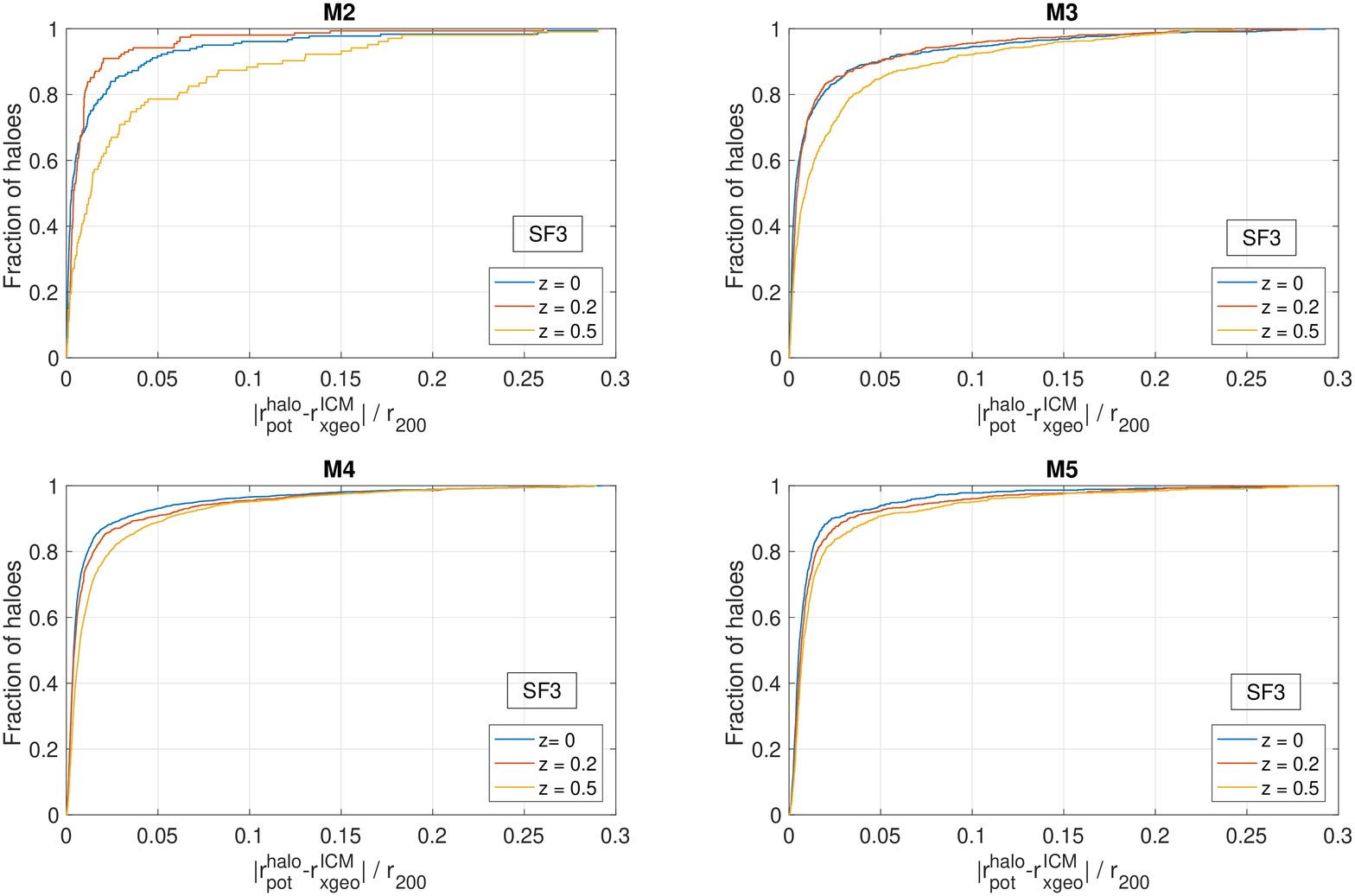}
\caption{Redshift evolution of the cumulative distributions of the rescaled PX offsets $r_{{\rm off,PX}}/r_{200}$ for different mass bins in the SF3 run. Lines of different colors refer to the results for different redshifts.}
\label{fig3}
\end{figure*}

\section{Results}
In this section, we analysis the results for the PB and PX offsets. To be concise, we mainly present results from the SF3 run in this section. Results from the other two runs are used for comparison and are only shown whenever necessary.

\subsection{Cumulative distribution of spatial offsets}
\label{section3.1}
As a matter of fact, the more massive halos tend to have larger physical separation $r_{\rm off}$ between baryonic and dark matter than the less massive ones. Moreover, a more massive halo would have a larger virial radius $r_{200}$, which could result in a larger $r_{\rm off}$ than less massive halo for the same level of unrelaxed-ness. To reduce the bias due to halo size, the rescaled centre offsets, i.e. $r_{\rm off}/r_{200}$ are probed. 

In Figure \ref{fig1a} and \ref{fig1b}, the cumulative distribution of the rescaled PB offsets $r_{\rm off, PB}/r_{200}$ and PX offsets $r_{\rm off,PX}/r_{200}$ for different mass bins for the SF3 run are respectively plotted. Two characteristics are notable. The first is that the cumulative distribution of the rescaled PB offsets $r_{\rm off, PB}/r_{200}$ as well as the PX offsets $r_{\rm off,PX}/r_{200}$ both show weak dependence on halo mass. The distribution curves for different mass bins almost overlap with each other.
The results for the mass bin M1 (green solid line) are contaminated by the statistical noise due to the deficiency of halo samples.

The second notable characteristic is that the PX offsets have a steeper distribution curve than the PB offsets. In other words, the PX offsets in most halo system are smaller than the PB offsets. 
Almost over 80 per cent of the halo systems have a PX offsets $r_{\rm off,PX}/r_{200}<0.05$ (equivalently $r_{\rm off,PX}\leq 10$ h$^{-1}$ kpc), while for the PB offsets this fraction is about 60$\%$. 
The result of PX offsets is consistent with that obtained by \citep{cui2016} from an SPH cosmological simulation which also include radiative cooling, star formation and kinetic feedback from supernovae with a softening length $\sim 7.5~{\rm h}^{-1}$ kpc (ignoring AGN feedback). Moreover, the level and distribution of PX offsets also agree with that reported in previous observation and simulation works \citep{Mann2012,Rossetti2016} despite the difference in the definition of centre offsets. 

In Figure \ref{fig2} and \ref{fig3}, the redshift evolution of the distribution of rescaled PB offsets $r_{\rm off, PB}/r_{200}$ and PX offsets $r_{\rm off,PX}/r_{200}$ for different mass bins in the SF3 run are respectively plotted. For each of the mass bin M2, M3, M4 and M5, the proportion of halos with a small offsets $r_{\rm off}/r_{200}$ (either $r_{\rm off, PB}/r_{200}$ or $r_{\rm off,PX}/r_{200}$) is larger for lower redshifts. 
This feature can be explained by the evolution of halos at low redshifts in the $\Lambda$CDM universe. The merge frequency slows down in the redshift range considered here, and hence the unrelaxed-ness decreases along with time. This conclusion holds it validity for both of the offsets definitions. As in Figure \ref{fig1}, the results for the mass bin M1 (top panels in Figure \ref{fig2} and \ref{fig3}) suffer from statistical noises since the bin is insufficiently sampled.

\begin{figure*}
\includegraphics[width=185mm]{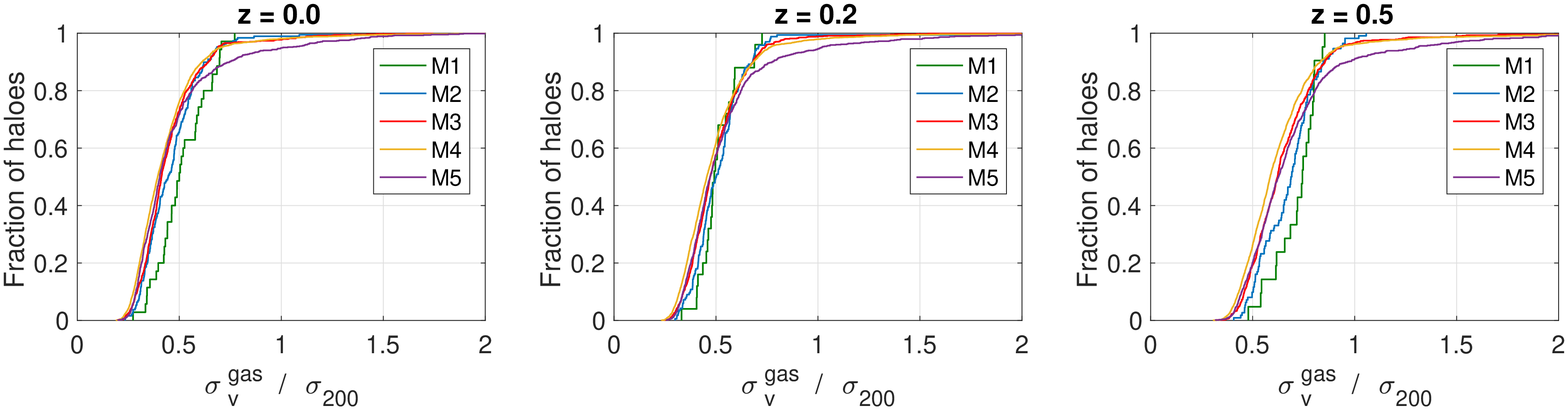}
\caption{Cumulative distributions of halos as a function of the rescaled velocity dispersion $\sigma_{v}^{\rm gas}/\sigma_{200}$ of the gas component at three different redshifts for the SF3 run. Lines of different colors refer to different halo mass bins as detailed in Table \ref{tab1}.}
\label{fig4}
\end{figure*}

\begin{figure*}
\includegraphics[width=70mm]{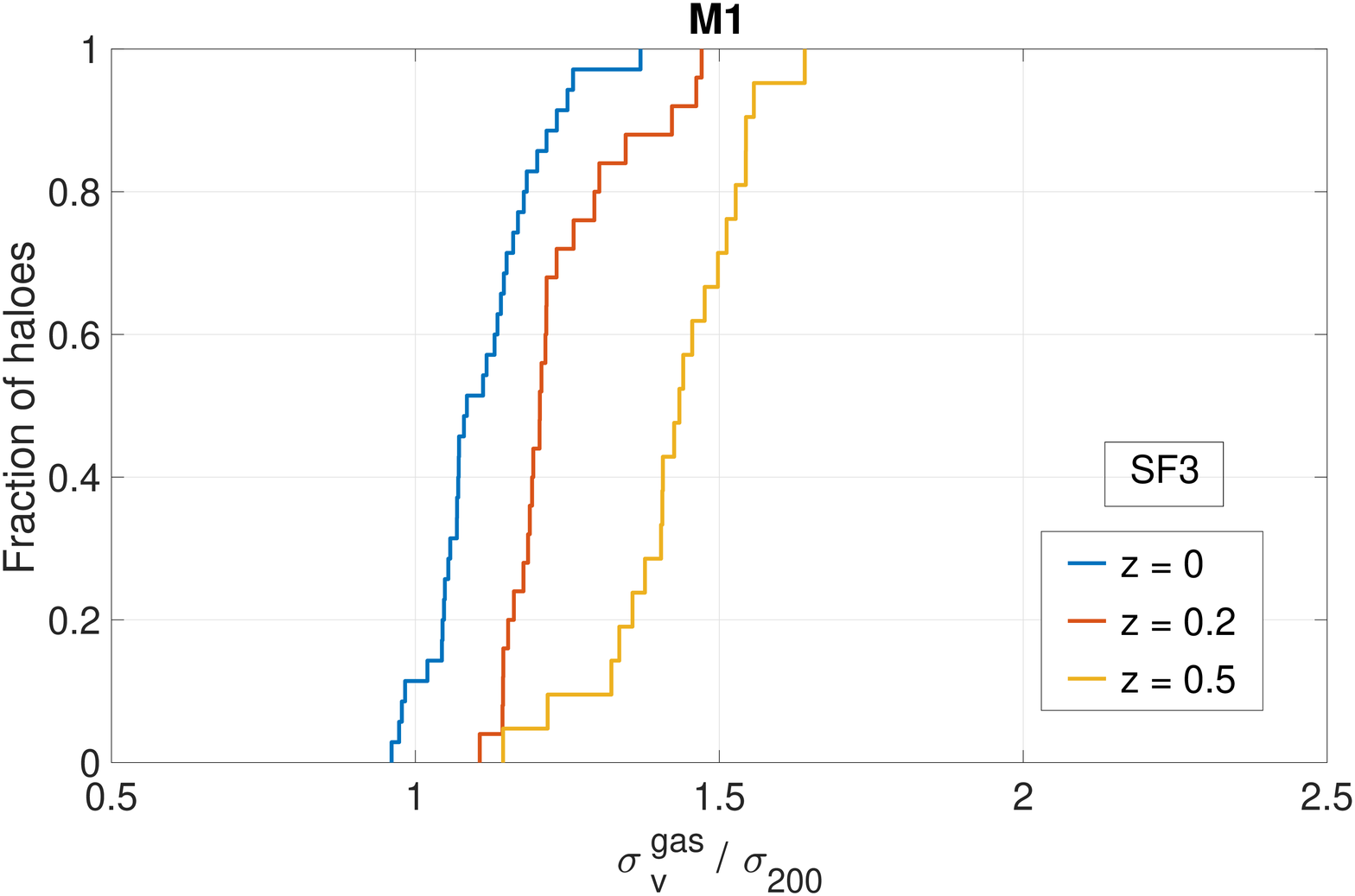}
\includegraphics[width=160mm]{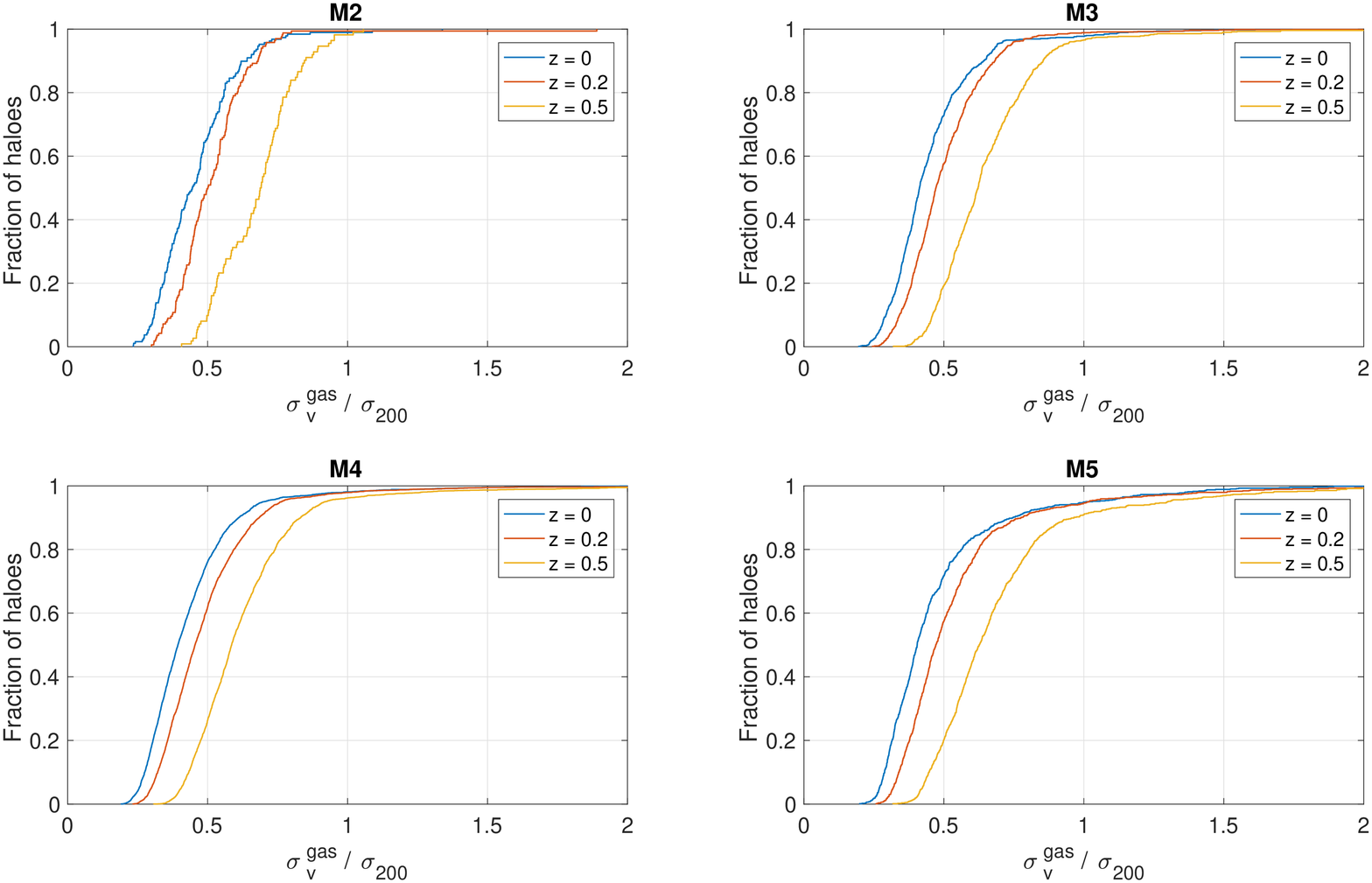}
\caption{Redshift evolution of the cumulative distributions of the rescaled velocity dispersion $\sigma_{v}^{\rm gas}/\sigma_{200}$ for halos in different mass bins in the SF3 run. Lines of different colors refer to the results for different redshifts.}
\label{fig5}
\end{figure*}

\begin{figure*}
\includegraphics[width=185mm]{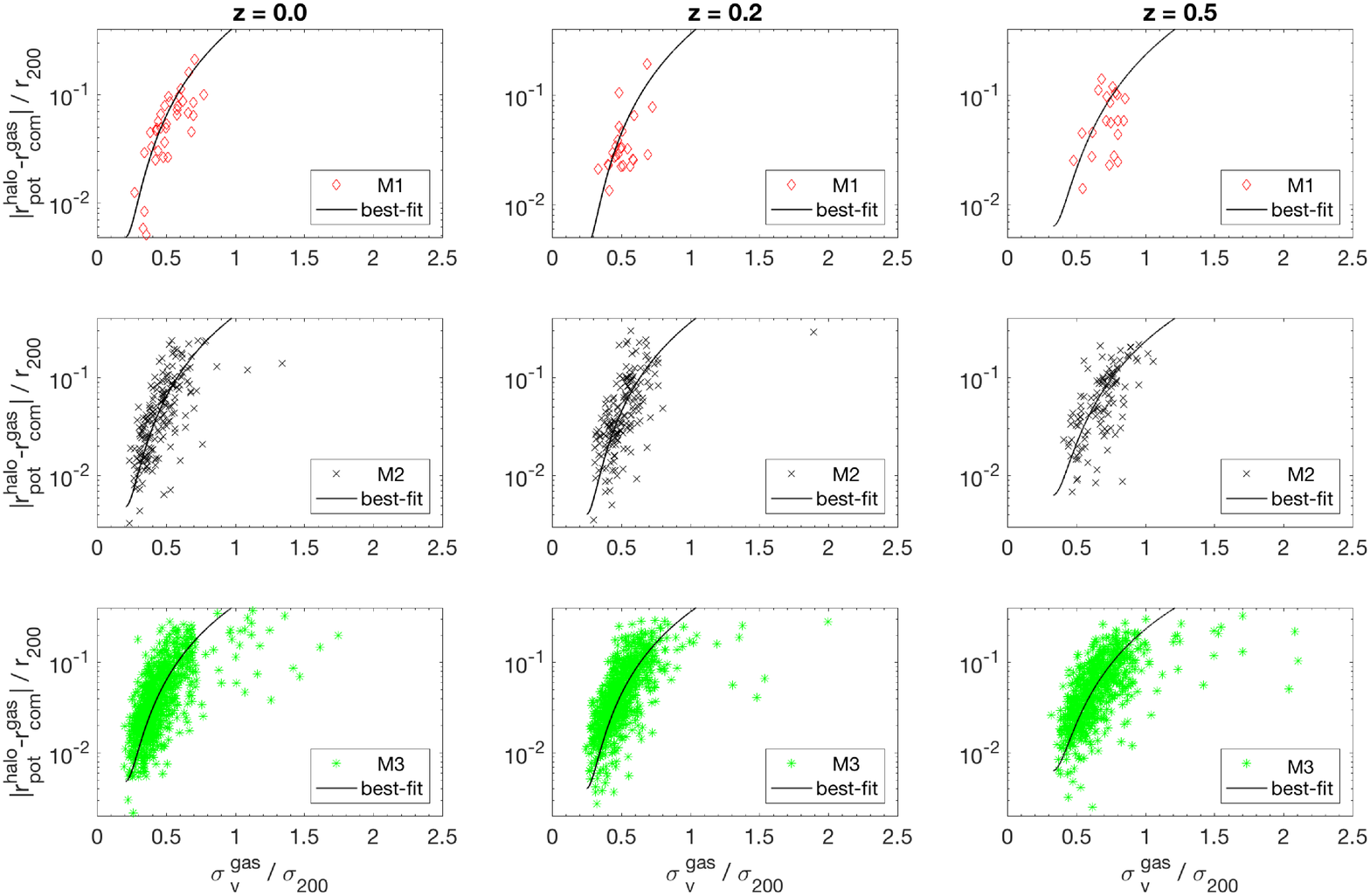}
\includegraphics[width=185mm]{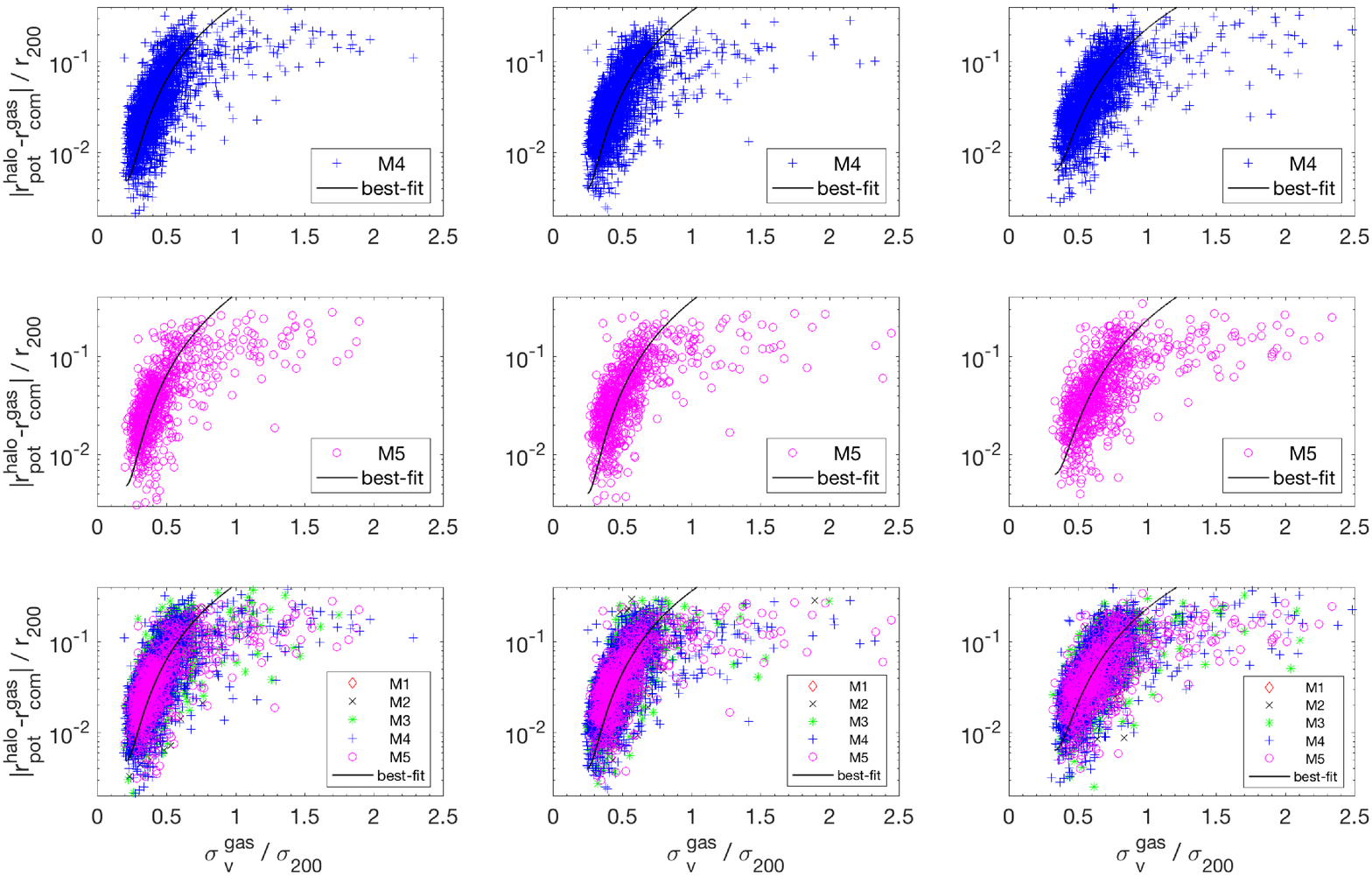}
\caption{Rescaled PB offsets, $r_{{\rm off,PB}}/r_{200}$, as a function of the rescaled gas velocity dispersion, $\sigma_{v}^{\rm gas}/\sigma_{200}$ for the SF3 run. Results for different mass bins at three different redshifts $z= 0, 0.2$, and $0.5$ are presented. Considering the large number of halos in the M5 bins, only a tenth of them are plotted above. The y-axis is presented in log scale. The black solid lines represent the quadratic function (\ref{pl1}) with the parameter values from the MCMC analysis of the entire halo catalogue (as presented in Table \ref{tab2}). }
\label{fig6}
\end{figure*}

\begin{figure*}
\includegraphics[width=185mm]{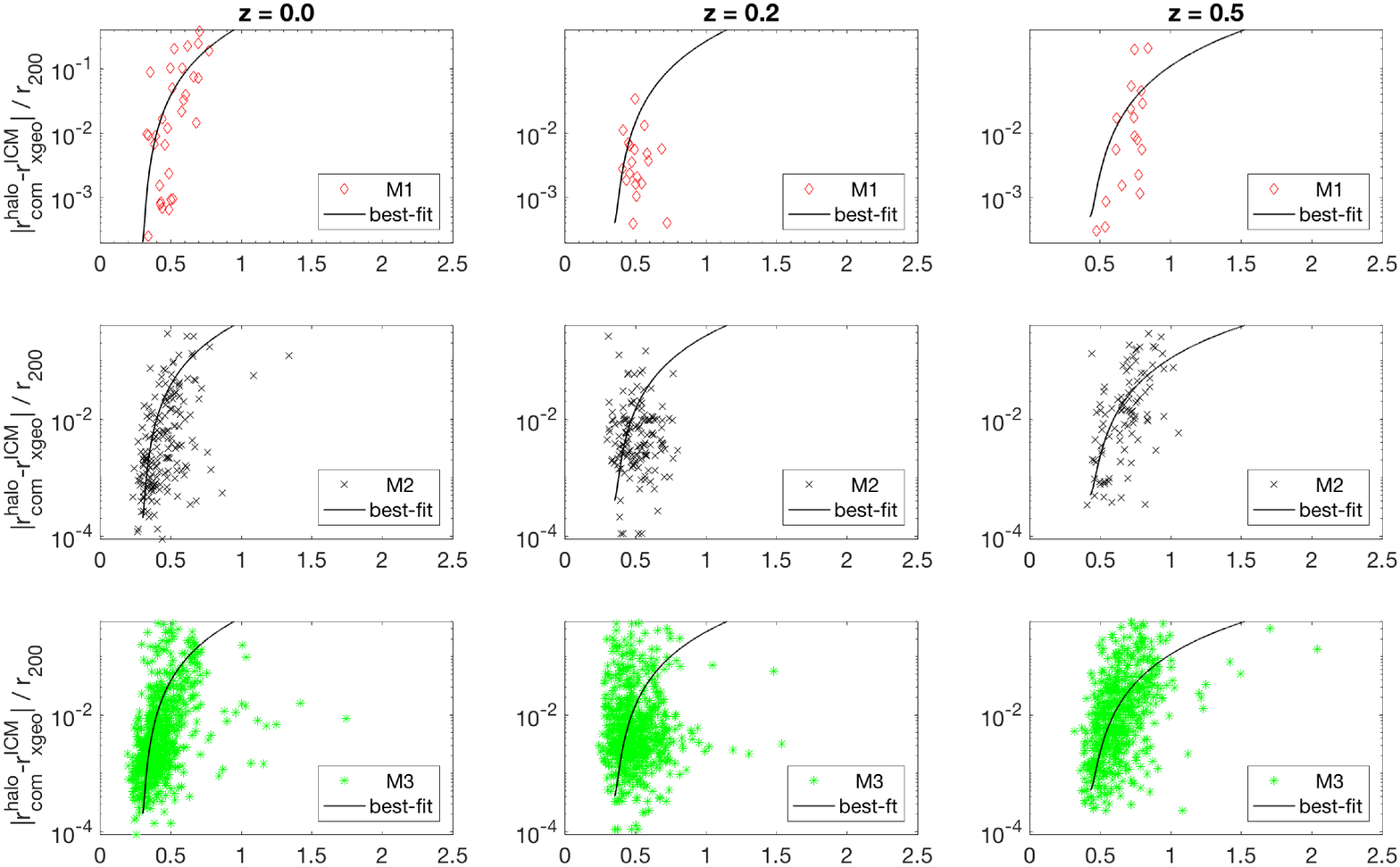}
\includegraphics[width=185mm]{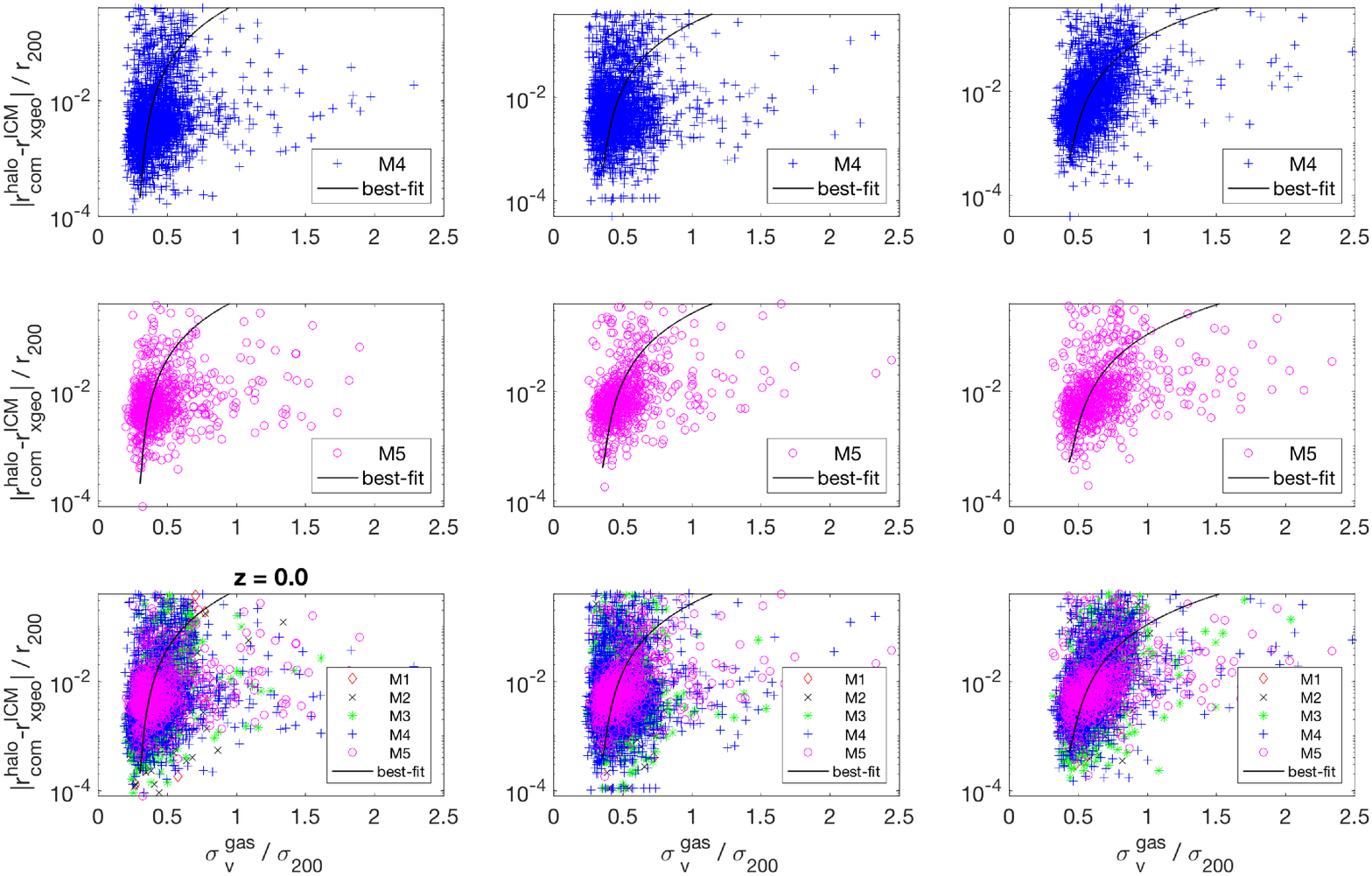}
\caption{Rescaled PX offsets, $r_{{\rm off,PX}}/r_{200}$, as a function of the rescaled gas velocity dispersion, $\sigma_{v}^{\rm gas}/\sigma_{200}$ for the SF3 run. Results for different mass bins at three different redshifts $z= 0, 0.2$, and $0.5$ are presented. Considering the large number of halos in the M5 bins, only a tenth of them are plotted above. The y-axis is presented in log scale. The black solid lines represent the quadratic function (\ref{pl1}) with the parameter values from the MCMC analysis of the entire halo catalogue (as presented in Table \ref{tab2}). }
\label{fig7}
\end{figure*}

\begin{figure*}
\includegraphics[width=185mm]{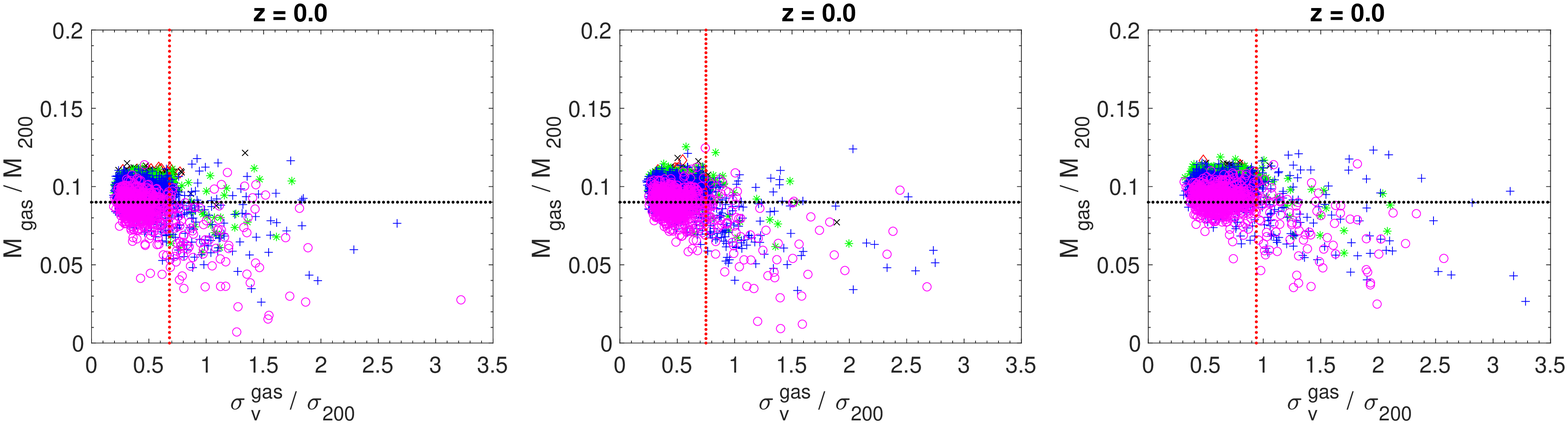}
\caption{Fraction of gas in the entire halo system as a function of the velocity dispersion of the gas component for the SF3 run. $M_{\rm gas}$ represents the virial mass of the gas component and $M_{\rm 200}$ is the mass of the whole system. The gas velocity dispersion $\sigma_{v}^{\rm gas}$ is rescaled by the velocity dispersion $\sigma_{200}$ for the whole system. Data points with different symbols (in the same patterns as used in Figure \ref{fig6} and \ref{fig7}) represent halos in the five different halo mass bins as detailed in Table \ref{tab1}. The results are presented at three different redshifts $z= 0, 0.2$, and $0.5$. The red dotted lines represents the threshold of $\sigma_{v}^{\rm gas}/\sigma_{\rm 200}$ below which the Bayesian analysis is applied. The $\sigma_{v}^{\rm gas}/\sigma_{\rm 200}$ thresholds are respectively 0.68 at $z=0$, 0.75 at $z=0.2$, 0.94 at $z=0.5$. The black solid lines represent the threshold of the gas fraction $M_{\rm gas}/M_{\rm 200}=0.09$. Considering the large number of halos in the M5 bins, only a tenth of them are plotted above.}
\label{fig8}
\end{figure*}

\subsection{Cumulative distribution of velocity dispersion}
\label{cdf}
We are interested in the correlation between rescaled spatial offsets $r_{\rm off}/r_{200}$ (either $r_{\rm off, PB}/r_{200}$ or $r_{\rm off,PX}/r_{200}$) and the dynamical state of the halo system. The dynamical state of a system, either dark matter or the gas component, can likely be reflected by its velocity dispersion, i.e., the root mean square (abbreviated as RMS or r.m.s) of the velocity variance of its corresponding particle members. To address our concern, we firstly study the distribution of three-dimension (3D) velocity dispersion of the gas component, dubbed as $\sigma_v^{\rm gas}$, in the halos. As the definition of $\mathbf{r}_{{\rm gas}}$, we also restrict to the gas that is residing within the halos.

To reduce the possible bias induced by the mass of the system, we introduce the rescaled velocity dispersion $\sigma_{v}^{\rm gas}/\sigma_{200}$, where the gas velocity dispersion $\sigma_{v}^{\rm gas}$ is rescaled by the 3D velocity dispersion $\sigma_{200}$ of the entire system. Following \citet{Munari2013}, $\sigma_{200}$ is given as
\be
\frac{\sigma_{200}}{{\rm km~s}^{-1}}=\frac{\sqrt{3}\sigma_{\rm 1D}}{{\rm km~s}^{-1}}=\sqrt{3} A_{\rm 1D}\left[\frac{h(z) M_{200}}{10^{15.0}\textmd{M}_\odot}\right]^{\alpha}, 
\label{sigma200}
\ee
where $A_{\rm 1D}\simeq 10^3$ and $\alpha \simeq 1/3$. $\sigma_{200}$ and $M_{200}$ are in unit of km s$^{-1}$ and h$^{-1}$ $\textmd{M}_\odot$ respectively. 

In Figure \ref{fig4}, the mass dependence of $\sigma_{v}^{\rm gas}/\sigma_{200}$ are shown. Like the rescaled PB and PX offsets, $\sigma_{v}^{\rm gas}/\sigma_{200}$ shows little dependence on the halo mass. The distribution curves of $\sigma_{v}^{\rm gas}/\sigma_{200}$ for different mass bins are in well agreement with each other within the statistical noise.

In Figure \ref{fig5}, the redshift evolution of the rescaled gas velocity dispersion $\sigma_{v}^{\rm gas}/\sigma_{200}$ are shown. Similar to that of the rescaled offsets shown in Figure \ref{fig2} and \ref{fig3}, the number of halos with a large $\sigma_{v}^{\rm gas}/\sigma_{200}$ decreases along with redshifts. It also implies that the merge rate and the un-relaxedness of the halos both go down with lower redshifts. It should also be noticed that the rescaled velocity dispersion $\sigma_{v}^{\rm gas}/\sigma_{200}$ depends more sensitively on the redshifts than the rescaled offsets. Thus, it can be taken as a proper indicator of the dynamic state of the system without mass bias.

\subsection{Correlation between centre offsets and velocity dispersion}
$r_{\rm off}/r_{200}$ and $\sigma_{v}^{\rm gas}/\sigma_{200}$ might can both be taken as the indicators of the dynamical state of the cluster system. Actually, the mass dependence and redshift evolution of  $r_{\rm off}/r_{200}$ and $\sigma_{v}^{\rm gas}/\sigma_{200}$ show evident similarity. Thus, it is interesting to study the correlation between them.

\subsubsection{$r_{\rm off}/r_{200}$ as a quadratic function of $\sigma_{v}^{\rm gas}/\sigma_{200}$}
\label{section3.2.3}
In Figure \ref{fig6} and \ref{fig7}, we respectively plot the PB offsets $r_{\rm off, PB}/r_{200}$ and PX offsets $r_{\rm off,PX}/r_{200}$ as a function of $\sigma_{v}^{\rm gas}/\sigma_{200}$ for halos in different mass bins and redshifts.
To quantitatively describe the correlation between the centre offsets $r_{\rm off}/r_{200}$ and the rescaled gas velocity dispersion $\sigma_{v}^{\rm gas}/\sigma_{200}$, we introduce the following quadratic function\footnote{The explanation for such a formalism is given and discussed in the final section.}, i.e.
{\be
\frac{r_{{\rm off}}}{r_{\rm 200}} = k_1  \left[\frac{\sigma_{v}^{\rm gas}}{\sigma_{200}} -k_2\right]^{2}+k_3,
\label{pl1}
\ee}
where $\sigma_v^{\rm gas}$ is the 3D velocity dispersion of the gas within the halo in the comoving coordinate system, and $\sigma_{200}$ is the 3D velocity dispersion of the entire halo, which is estimated by equation (\ref{sigma200}). $r_{\rm off}/r_{200}$ can be either the PB offsets $r_{\rm off, PB}/r_{200}$ or the PX offsets $r_{\rm off,PX}/r_{200}$.

The model parameter vector is defined as $\mathbf{\lambda} = (k_1, k_2, k_3)$. We apply Bayesian inference techniques to determine the posterior probability distribution $P(\mathbf{\lambda}|\mathcal{D})$ of the model parameters, given the simulation data $\mathcal{D}$. We sample the posterior distribution $P(\mathbf{\lambda}|\mathcal{D})$ using the Markov-Chain Monte-Carlo (MCMC) method. The $r_{{\rm off}}/r_{200}$ for each halo is weighted by $w_{\rm off}=|(\sigma_{v}^{\rm gas}/\sigma_{200})\cdot 4.5\times 10^{-2}|$ in the statistical analysis in order to obtain a mean $\chi_{\rm min}^2 \sim 1$.
The halo samples from the SF1, SF2 and SF3 simulations are all explored in our statistical study. For each simulation, we perform our statistical analysis on the whole halo sample in each snapshot, i.e., at $z = 0$, $0.2$ and $0.5$ respectively to obtain the value of the model parameters at different redshifts. The black lines in Figure \ref{fig6} and \ref{fig7} indicate the fitting results. The results for the three simulations are similar so that we mainly present those for the SF3. The analyses are restricted to those systems with $\sigma_{v}^{\rm gas}/\sigma_{200}\leq 0.68, 0.75$, and $0.94$ (see Figure \ref{fig8}).

\begin{table*}
\begin{tabular}{|c|c|c|c|c|c|c|c|c|}
\hline
 & \multicolumn{4}{|c|}{PB offsets (SF3, $1\sigma$ C.L.)} & \multicolumn{4}{|c|}{PX offsets (SF3 , $1\sigma$ C.L.)} \\ 
\hline
(1) & (2) & (3) & (4) & (5) & (6) & (7) & (8) & (9) \\ 
\hline
redshift & $z=0$ & $z=0.2$ & $z=0.5$ & {\bf Mean} & $z=0$ & $z=0.2$ & $z=0.5$ & {\bf Mean} \\ 
\hline
\hline
k1        & $0.667_{-0.294}^{+0.396}$ & $0.626_{-0.377}^{+0.398}$ & $0.502_{-0.340}^{+0.422}$  &  ${\bf 0.599_{-0.337}^{+0.405}}$    & $0.945_{-0.264}^{+0.394}$ & $0.626_{-0.250}^{+0.485}$ & $0.331_{-0.286}^{+0.445}$  &   ${\bf 0.634_{-0.267}^{+0.441}}$ \\
\hline
k2  & $0.205_{-0.048}^{+0.051}$ & $0.247_{-0.034}^{+0.044}$ & $0.330_{-0.032}^{+0.061}$  &   ${\bf 0.261_{-0.038}^{+0.052}}$       &  $0.302_{-0.031}^{+0.045}$ & $0.352_{-0.021}^{+0.059}$ & $0.427_{-0.041}^{+0.057}$  &    ${\bf 0.361_{-0.031}^{+0.054}}$       \\
\hline
k3       & $0.005_{-0.004}^{+0.005}$ & $0.004_{-0.004}^{+0.005}$ & $0.006_{-0.006}^{+0.007}$  &    ${\bf 0.005_{-0.005}^{+0.006}}$   &  $0.0002_{-0.0001}^{+0.0002}$ & $0.0004_{-0.0003}^{+0.0005}$ & $0.0005_{-0.0001}^{+0.0004}$ &    ${\bf 0.0004_{-0.0002}^{+0.0003}}$      \\
\hline
\end{tabular}
\caption{Marginalized values of the parameter $k_1$, $k_2$ and $k_3$ from the MCMC analysis of the whole sample for both the PB and PX offsets for the SF3 run. The (2) to (4) columns list the universal marginalized mean values of $k_1$ and $k_2$ with $68$ per cent confidence intervals at different redshifts for the PB offsets, while the (6) to (8) columns list those for the PX offsets. The (5) and (9) columns list the averaged values from (2) to (4) and from (6) to (8) columns, respectively.}
\label{tab2}
\end{table*}

\subsubsection{parameters $k_1$, $k_2$ and $k_3$}
The marginalized values of $k_1$, $k_2$ and $k_3$ from the MCMC analysis of the halos at different redshifts for two types of offsets definitions are presented in Table \ref{tab2}. 
The conclusions for the rescaled offsets $r_{\rm off}/r_{200}$ in this subsection are valid for both of the PB offsets $r_{\rm off, PB}/r_{200}$ and PX offsets $r_{\rm off,PX}/r_{200}$.

The parameter $k_1$ represents the amplitude of the spatial offsets $r_{\rm off}/r_{200}$. It to a certain degree describes that how strong $r_{\rm off}/r_{200}$ 
correlates with $\sigma_{v}^{\rm gas}/\sigma_{200}$. As shown in Table \ref{tab2}, for both types of $r_{\rm off}/r_{200}$, the simulation data gives a non-zero $k_1$, while the relatively large confidence regions of $k_1$ (up to 30 $\sim$ 40 per cent of the mean value) reflects that the constraint is loose. For certain redshift, the PB and PX offsets have a $k_1$ that are coincident with each other at $1\sigma$ confidence level.

The parameter $k_2$ is related to the dynamical state of the halo system. According to the equation (\ref{pl1}), the spatial offsets $r_{\rm off}/r_{200}$ obtains its minimum value when $\sigma_{v}^{\rm gas}/\sigma_{\rm 200}=k_2$. Physically, the spatial offsets $r_{\rm off}/r_{200}$ for a system is believed to have its minimum value when the system is dynamically stable or relaxed. Therefore, $k_2$ is equivalently the critical value of $\sigma_{v}^{\rm gas}/\sigma_{\rm 200}$ that a system can have to its maximum stability. For both types of offsets, the marginalized value of $k_2$ shows a more apparent redshift evolution than that for the parameter $k_1$, in analogy to what we found from comparing the redshift evolution of the distribution of $r_{{\rm off}}/r_{200}$ and $\sigma_{v}^{\rm gas}/\sigma_{\rm 200}$ in section \ref{cdf}. The systems at higher redshift tend to have a larger $k_2$. Given the same redshift, the marginalized value of $k_2$ for the PX offsets is relatively larger than that for the PB offsets.

The parameter $k_3$ represents the residuals of the rescaled offsets $r_{\rm off}/r_{200}$ even though the system reaches its maximum stability at $\sigma_{v}^{\rm gas}/\sigma_{\rm 200}=k_2$. For both types of offsets, the marginalized values $k_3$ for different redshifts are consistent with each other within the statistical uncertainties. The small magnitude of the $k_3$ term are in fact in agreement with the magnitudes of the spacial centres offsets in relaxed clusters reported in the literatures.

\section{Conclusions and Discussion}
In this paper, the centre offsets between the ICM gas and the entire halo system are studied using three hydrodynamical cosmological simulations. We focus on two kinds of centre offsets: one is the three-dimensional PB offsets between the gravitational potential minimum of the entire halo and the barycentre of the ICM, and the other is the two-dimensional PX offsets between the potential minimum of the halo and the iterative centroid of the projected synthetic X-ray emission of the halo. 
The halos at higher redshifts tend to have a slightly larger centre offsets than those at lower redshifts (see Figure \ref{fig2}). This possibly reflects the fact that the merge frequency 
and un-relaxedness of a galaxy cluster goes down with the passage of time. We further probe the gas velocity dispersion. The cumulative distribution of $\sigma_v^{\rm gas}/\sigma_{200}$ shows that the halos at higher redshifts tend to have an apparently larger $\sigma_v^{\rm gas}/\sigma_{200}$ than those at lower redshifts. 
Nevertheless, both the $\sigma_v^{\rm gas}/\sigma_{200}$ and $r_{\rm off}/r_{200}$ shows weak dependence on halo mass and the resolution. Therefore, both of them can be used as indicators of the dynamic state of galaxy clusters with different virial mass.  In fact, the offsets has been used as an criterion to select relaxed cluster in previous work (e.g., \citealt{cui2017}), despite the difference on the definition of offsets. For both types of offsets, we find that the correlation between the rescaled centre offsets $r_{\rm off}/r_{200}$ and the rescaled 3D gas velocity dispersion, $\sigma_v^{\rm gas}/\sigma_{200}$ could be approximately described by a quadratic function as equation (\ref{pl1}). However, the correlation of $r_{\rm off, PX}/r_{200}$ is relatively weaker.  A Bayesian analysis with MCMC method was employed to estimate the model parameters $k_1$, $k_2$ and $k_3$. On the other hand, the magnitude of $\sigma_v^{\rm gas}/\sigma_{200}$ is affected by the gas mass fraction of the halos. We define the gas mass fraction as $f_{\rm gas} \equiv M_{\rm gas}/M_{200}$. The gas fraction $f_{\rm gas}$ as a function of the velocity dispersion $\sigma_{v}^{\rm gas}/\sigma_{200}$ is plotted in Figure \ref{fig8}. One can see that some halos with low $f_{\rm gas}$ tend to have large $\sigma_{v}^{\rm gas}/\sigma_{200}$, thus might be dynamically more un-relaxed than those with high $f_{\rm gas}$, and might weaken the correlation between rescaled offsets and velocity dispersion.

Some comments are necessary on the quadratic function (\ref{pl1}). The function (\ref{pl1}) we introduced can be roughly considered to be an interference of the virial theorem. For a collapsed system with virial radius $r_{200}$ and mass $M$, the virial theorem has the form of $2T+U=0$, where $U=-\alpha GM^2/r_{200}$ and $T=\frac{1}{2}\beta M \langle v^2 \rangle$. $\alpha$ and $\beta$ are constants, and their values are dependent on the specific profile of the system. $\langle v^2 \rangle$ is the variation of the velocity of particles within, which equals the 3D velocity dispersion $\sigma_{\rm v}^2$ assuming a Gaussian velocity distribution.
The deviation of the potential energy from equilibrium state, which is related to the spatial centre offsets of the entire system, $\Delta r \equiv r_{\rm off}$, and can be written as 
\be
\Delta U \sim \alpha \frac{GM^2}{r_{200}}\cdot \frac{\Delta r}{r_{200}} \sim \alpha \frac{GM^2}{r_{200}}\cdot \frac{r_{\rm off}}{r_{200}}, 
\label{virialV}
\ee

Assuming that the characteristic gas velocity dispersion is $\sigma_{v, {\rm relaxed}}^{\rm gas} =k_2 \sigma_{\rm 200}$ in a relaxed system, the corresponding perturbation of the kinematic energy of the gas component can be expressed as 
\be
\begin{aligned}
\Delta T_{\rm gas} \sim \frac{1}{2}\beta M_{\rm gas} (\sigma_{v}^{\rm gas} -k_2 \sigma_{\rm 200})^2  \\
\sim \frac{1}{2}\beta M_{\rm gas} \sigma_{\rm 200}^2  \cdot (\sigma_{v}^{\rm gas}/\sigma_{200} - k_2)^{2} \\
\sim f_{\rm gas} T \cdot (\sigma_{v}^{\rm gas}/\sigma_{200} - k_2)^{2}. 
\end{aligned}
\label{virialE}
\ee
We further assume that the perturbation of the kinematic energy of the dark matter is a few times of that of gas, i.e., $\Delta T_{\rm cdm} \sim \gamma \Delta T_{\rm gas}$. 
Therefore,  $\Delta T = \Delta T_{\rm cdm}+\Delta T_{\rm gas} \sim (1+\gamma) \Delta T_{\rm gas} \sim (1+\gamma) f_{\rm gas} T \cdot (\sigma_{v}^{\rm gas}/\sigma_{200} - k_2)^{2}$. Substituting these terms into $\Delta T / T \approx \Delta U/U$, one obtains 
\be
r_{\rm off}/r_{200} \propto  (\sigma_{v}^{\rm gas}/\sigma_{200} - k_2)^{2}.
\ee

It should be emphasized that the model was somewhat oversimplified. For instance, the thermal energy of the gas, which could be an important factor, was not included in our derivation. Recent works using high resolution AMR simulations showed that the thermal energy is of almost the same order of magnitude as the kinematic energy (due to turbulent motions, the ratio is 2-5) of the ICM (\citet{vazza2011}, \citet{Schmidt2017}, and references therein).
Considering this, the perturbation of thermal energy of gas would generally have the same order of magnitude as $\Delta T_{\rm gas}$. This could be accommodated by a factor in equation (\ref{pl1}) and would not change the form of the equation significantly. Meanwhile, these studies also showed that the turbulent motions of ICM remain non-negligible even in relaxed clusters. Hence, a non-zero $k_2$ is expected.

More comments are also necessary for the PX offsets, and its correlation with the velocity dispersion.
One factor is the projection effect. The PB offsets is calculated directly from the 3D simulation data and thus is one of the intrinsic physical separation between the two centres. It is closely related to the inner kinematic or dynamic processes of the system. The PX offsets is a 2D offsets between the potential minimum of the halo and the iterative centroid of the projected synthetic X-ray emission of the halo. To calculate the PX offsets, one need to first simulate the X-ray emission of the halo due to the metallicity and the temperature of the gas. Then the synthetic X-ray photons are projected along the line-of-sight (l.o.s) direction to obtain the 2D mock X-ray profile. Any structures along the l.o.s would introduce contamination to the results of PX offsets. About $10$ per cent of the halos in our samples suffer from this effect. These halos usually have an abnormally large PX offsets ($r_{\rm off,PX}/r_{200} >0.3$) but a reasonable PB offsets ($r_{\rm off, PB}/r_{200} <0.3$, which implies that the system is actually dynamically relaxed). We examined these halos and found that over $98\%$ of these halos have significant X-ray contamination from other structures within the projected distance of the virial radius along the l.o.s direction. These contaminated halos were not included into our statistical analysis.

Moreover, the model used to generate the X-ray photons would also introduce biases to the results. In generating the 2D mock X-ray map of the halos that are employed for numerical studies in this paper, we do not convolve the X-ray photons with the responding characteristics of a specific instrument, defined by the redistribution matrix file (RMF) and the ancillary response file (ARF) of any specific instrument, which could be another source of biases. The correlation between the PX offsets and the $\sigma_{v}^{\rm gas}/\sigma_{200}$ could possibly have been weakened by these biases.

Despite its large scatter, the correlation we obtained in this paper provides an alternative way of characterizing the centre offsets for clusters, other than the statistical approaches adopted by \citet{john07b}, \citet{More2015}, \citet{viola2015}, and references therein. Our work shows the possibility of establishing empirical relations between some physical properties of the galaxy clusters and groups, and centre offsets. These correlations, if properly calibrated, could be employed in the future data analysis as an alternative approach to determine the `true' centre of the clusters of galaxies from the observations.

Our investigation also shows that the intrinsic physical separation of gas centre is significantly larger than the two-dimensional PX offsets. The offsets of BCGs from the minimum of potential well is generally comparable or smaller than the PX offsets (e.g., \citealt{cui2016}). Considerable physical separation between gas centre and BCG is expected, which might can partly explain the finding in \cite{Sehgal2013}, i.e., a lower recovered SZ signal than Planck.

Last but not the least, it should be noticed that the barycentre or the X-ray centroid of the ICM or IGM do not always coincide with the BCGs, the latter of which are often considered as the cluster centre in the galaxy and cluster observations as well as in mass reconstruction by stacked weak lensing. Further studies are urged to investigate their possible relations and its impact on the centre identification and mass estimation for galaxy groups or clusters, as well as on the properties of SZ selected clusters. Related research are undertaking.

\section*{Acknowledgments}
We would like to thank the anonymous referee for their informative comments and constructive suggestions in improving the manuscript.
We would like to express our gratitude to Dr. Wei-Peng Lin for providing us the simulation data of SF1 and SF2 for the research. We would also like to thank Dr. Veronica Biffi for providing the PHOX code and the permission to use it for the synthetic X-ray profile generation in this work.
We are grateful to the stimulating discussions with Dr. Wei-Peng Lin, Dr. Yang Wang and Dr. Shi-Hong Liao. The computations in this paper was supported partly by the HPC facilities at SYSU. 
We also thank Dr. Zhi-Qi Huang, Dr. Yi-Jung Yang, and Dr. Fu-Peng Zhang for their insightful comments on this work.
This work is supported by the National Key Program for Science and Technology Research and Development (2017YFB0203300) and the National Natural Science Foundation of China (NFSC) through grant 11733010.
WSZ acknowledges support from the the National Natural Science Foundation of China (NSFC) under grants 11673077 and the Fundamental Research Funds for the Central Universities.

\appendix




\begin{thebibliography}{}
\bibitem[\protect\citeauthoryear{Ade et al.}{2016}]{Planck2015}
Ade, P. A. R. et al. [Planck Collaboration] 2016, A\&A, 594, A13

\bibitem[\protect\citeauthoryear{Andreon \& Moretti}{2011}]{AM2011}
Andreon, S. \& Moretti, A. 2011, A\&A, 536, A37

\bibitem[\protect\citeauthoryear{Berlind et al.}{2003}]{Berlind2003}
Berlind, A. A., Weinberg, D. H., Benson, A. J., Baugh, C. M., Cole, S., et al., 2003, ApJ, 593, 1

\bibitem[\protect\citeauthoryear{Biesiadzinski et al.}{2012}]{Bie2012}
Biesiadzinski, T., Mcmahon, J., Miller, C. J., Nord, B., \& Shaw, L. 2012, ApJ, 757, 1


\bibitem[\protect\citeauthoryear{Biffi et al.}{2012}]{Biffi2012}
Biffi V., Dolag K., B\"{o}hringer H., \& Lemson G. 2012, MNRAS, 420, 3545

\bibitem[\protect\citeauthoryear{Biffi et al.}{2012}]{Biffi2013}
Biffi V., Dolag K., \& B\"{o}hringer H. 2013, MNRAS, 428, 1395


\bibitem[\protect\citeauthoryear{B\"{o}hringer et al.}{2010}]{Bohringer2010}
B\"{o}hringer H., et al. 2010, A\&A, 514, A32

 \bibitem[\protect\citeauthoryear{Brada\v{c} et al.}{2006}]{Bradac2006}
Brada\v{c} M., Clowe D., Gonzalez A. H. et al. 2006, ApJ, 652, 937

\bibitem[\protect\citeauthoryear{Clowe et al.}{2006}]{Clowe2006}
Clowe D., Brada\v{c} M., Gonzalez A. H., Markevitch M., Randall S. W., Jones C., Zaritsky D. 2006, ApJ, 648, L109

\bibitem[\protect\citeauthoryear{Cui et al.}{2016}]{cui2016}
Cui W. G., Power C., Biffi V. et al. 2016, MNRAS, 456, 2566

\bibitem[\protect\citeauthoryear{Cui et al.}{2017}]{cui2017}
Cui W. G., Power C., Borgani S., Knebe A., Lewis G. F., Murante G., \& Poole G. B. 2017, MNRAS, 464, 2502


\bibitem[\protect\citeauthoryear{Dietrich et al.}{2012}]{die2012}
Dietrich J. P., BÂhnert A., Lombardi M., Hilbert S., \& Hartlap J. 2012, MNRAS, 419, 3547

\bibitem[\protect\citeauthoryear{Dong et al.}{2014}]{Dong2014}
Dong X. C., Lin W. P., Kang X., Ocean Wang, Yang, Dutton Aaron A., Macci ograve, \& Andrea V., 2014,  ApJL,  791,  L33 


 \bibitem[\protect\citeauthoryear{Du \& Fan}{2014}]{DF2014}
Du W., Fan Z., 2014, ApJ, 785, 57

 \bibitem[\protect\citeauthoryear{Evrard et al.}{2008}]{Evrard2008}
Evrard A. E. et al., 2008, ApJ, 672, 122

 \bibitem[\protect\citeauthoryear{Gao \& White}{2006}]{GW2006}
Gao L., White S. D. M. 2006, MNRAS, 373, 65

\bibitem[\protect\citeauthoryear{George et al.}{2012}]{george2012}
George M. R., Leauthaud A., Bundy K., et al. 2012, ApJ, 757, 2

\bibitem[\protect\citeauthoryear{Gupta et al.}{2017}]{Gupta2017}
Gupta, N., Saro, A., Mohr, J. J., Dolag, K., \& Liu, J., 2017, MNRAS, 469, 3069

\bibitem[\protect\citeauthoryear{Hironao et al.}{2015}]{hi2015}
Hironao M., Surhud M., Mandelbaum R., et al. 2015, ApJ, 806, 1

\bibitem[\protect\citeauthoryear{Hudson et al.}{2010}]{Hudson2010}
Hudson, D. S.; Mittal, R.; Reiprich, T. H.; Nulsen, P. E. J.; Andernach, H.; Sarazin, C. L., 2010, A\&A, 513, 37

 \bibitem[\protect\citeauthoryear{Johnston et al.}{2007a}]{john07a}
Johnston D. E., Sheldon E. S., Tasitsiomi A., et al. 2007a, ApJ, 656, 27

 \bibitem[\protect\citeauthoryear{Johnston et al.}{2007b}]{john07b}
Johnston D. E., Sheldon E. S., Wechsler R. H., et al. 2007b, arXiv:0709.1159

 \bibitem[\protect\citeauthoryear{Kim et al.}{2017}]{kim2017}
Kim, S., Y., Peter, A. H. G., \& Wittman, D. 2017, MNRAS, 469, 1414

\bibitem[\protect\citeauthoryear{Knollmann \& Knebe}{2009}]{KK2009}
Knollmann S. R., \& Knebe A., 2009, ApJS, 182, 608

\bibitem[\protect\citeauthoryear{Liao et al.}{2016}]{Liao2016}
Liao S., Gao L., Frenk C. S., Guo Q., \& Wang J. 2016, submitted to MNRAS, arXiv: 1610.07592v1

\bibitem[\protect\citeauthoryear{Lin \& Mohr}{2004}]{Lin2004}
Lin, Y-. T., \& Mohr, J. J., 2004, ApJ, 617, 879

\bibitem[\protect\citeauthoryear{Mandelbaum et al.}{2010}]{mandelbaum2010}
Mandelbaum R., Seljak U., Baldauf T., \& Smith R. E. 2010, MNRAS, 405, 2078

\bibitem[\protect\citeauthoryear{Mann \& Ebeling}{2012}]{Mann2012}
Mann, A. W., Ebeling, H., 2012, MNRAS, 420, 2120

\bibitem[\protect\citeauthoryear{Mantz et al.}{2015}]{Mantz2015}
Mantz A. B., Allen S. W., Morris R. G., Schmidt R. W., von der Linden A., \& Urban O. 2015, MNRAS, 449, 199

\bibitem[\protect\citeauthoryear{More et al.}{2015}]{More2015}
More S., Miyatake H., Mandelbaum R. et al. 2015, ApJ, 806, 2

\bibitem[\protect\citeauthoryear{Morrison \& McCammon}{1983}]{MM1983}
Morrison R. \& McCammon D. 1983, ApJ, 270, 119

\bibitem[\protect\citeauthoryear{Munari et al.}{2013}]{Munari2013}
Munari E., Biviano A., Borgani S., Murante G. \& Fabjan D. 2013,  MNRAS, 430, 2638

\bibitem[\protect\citeauthoryear{Navarro et al.}{1996}]{NFW1996}
Navarro J. F., Frenk C. S., \& White S. D. M. 1996, ApJ, 462, 563

\bibitem[\protect\citeauthoryear{Navarro et al.}{1997}]{NFW1997}
Navarro J. F., Frenk C. S., \& White S. D. M. 1997, ApJ, 490, 493

\bibitem[\protect\citeauthoryear{Oguri et al.}{2010}]{oguri2010}
Oguri M., Takada M., Okabe N., Smith G. P., 2010, MNRAS, 405, 2215


\bibitem[\protect\citeauthoryear{Rasia et al.}{2013}]{Rasia2013}
Rasia E., Meneghetti M. \& Ettori S. 2013, The Astronomical Review, 8, 40

\bibitem[\protect\citeauthoryear{Robertson et al.}{2017}]{robertson2017}
Robertson A., Massey R., \& Eke V. 2017, MNRAS, 465, 569


\bibitem[\protect\citeauthoryear{Rossetti et al.}{2016}]{Rossetti2016}
Rossetti, M., Gastaldello, F., Ferioli, G., Bersanelli, M., De Grandi, S., et al., 2016, MNRAS, 457, 4515

\bibitem[\protect\citeauthoryear{Rozo et al.}{2011}]{rozo2011}
Rozo E., Rykoff E., Koester B., et al. 2011, ApJ, 740, 53

\bibitem[\protect\citeauthoryear{Sanderson et al.}{2009}]{Sanderson2009}
Sanderson, A. J. R., Edge, A. C., \& Smith, G. P., 2009, MNRAS, 398, 1698

\bibitem[\protect\citeauthoryear{Saro et al.}{2015}]{Saro2015}
Saro, A., Bocquet, S., Rozo, E., Benson, B. A., Mohr, J., Rykoff, E. S., et al., 2015, MNRAS, 454, 2305

\bibitem[\protect\citeauthoryear{Schmidt et al.}{2017}]{Schmidt2017}	
Schmidt, W., Byrohl, C., Engels, J. F., Behrens, C., \& Niemeyer, J. C., 2017, MNRAS, 470, 142

\bibitem[\protect\citeauthoryear{Sehgal et al.}{2013}]{Sehgal2013}
Sehgal N. et al., 2013, ApJ, 767, 38

\bibitem[\protect\citeauthoryear{Sereno et al.}{2015}]{sereno2015}
Sereno M., Veropalumbo A., Marulli F., Covone G., Moscardini L., Cimatti A. 2015, MNRAS, 449, 4147

 \bibitem[\protect\citeauthoryear{Shan et al.}{2010}]{Shan2010}
Shan H.-Y., Qin B., Fort B., Tao C., Wu X.-P., \& Zhao H.-S. 2010, MNRAS, 406, 1134

\bibitem[\protect\citeauthoryear{Smith et al.}{2001}]{Smith2001}
Smith R. K., Brickhouse N. S., Liedahl D. A., \& Raymond J. C. 2001, ApJ, 556, L91


\bibitem[\protect\citeauthoryear{Skibba et al.}{2011}]{skibba2011}
Skibba R. A. \& Macci${\rm \grave{o}}$, A. V. 2011, MNRAS, 416, 2388

\bibitem[\protect\citeauthoryear{Springel}{2005}]{springel2005}
Springel V., 2005, MNRAS, 364, 1105

\bibitem[\protect\citeauthoryear{Surhud et al.}{2015}]{sur2015}
Surhud M., Hironao M., Mandelbaum R., et al. 2015, ApJ, 806, 2

\bibitem[\protect\citeauthoryear{Tang et al.}{2018}]{Tang2018}
Tang L., Lin W. P., Cui W. G., Kang Xi, Wang Y., Contini E., \& Yu, Y. 2018, arXiv:1804.03335


 \bibitem[\protect\citeauthoryear{van den Bosch et al.}{2013}]{van2013}
van den Bosch F. C., More S., Cacciato M., Mo H., \& Yang X. 2013, MNRAS, 430, 725

\bibitem[\protect\citeauthoryear{van den Bosch et al.}{2005}]{van2005}
van den Bosch F. C., Weinmann S. M., Yang X., Mo H. J., Li C., Jing Y. P., 2005, MNRAS, 361, 1203

\bibitem[\protect\citeauthoryear{Vazza et al.}{2011}]{vazza2011}
Vazza, F., Brunetti, G., Gheller, C., Brunino, R., \& Br\"{u}ggen, M.,  2011, A\&A, 529, 17

\bibitem[\protect\citeauthoryear{Viola et al.}{2015}]{viola2015}
Viola M., Cacciato M., Brouwer M., et al. 2015, MNRAS, 452, 3529

\bibitem[\protect\citeauthoryear{Wang et al.}{2014}]{Wang2014}
Wang Y. O., Lin W. P., Kang X., Dutton A., Yu Y., Macci O., \& Andrea V., 2014, ApJ, 786, 8 


\bibitem[\protect\citeauthoryear{Wuyts et al.}{2012}]{wuyts2012}
Wuyts S., Schreiber N. M. F., Genzel R., et al. 2012, ApJ, 753, 114

\bibitem[\protect\citeauthoryear{Zitrin et al.}{2012}]{zitrin2012}
Zitrin A., Bartelmann M., Umetsu K., Oguri M., \& Broadhurst T., 2012, MNRAS, 426, 2944

\bibitem[\protect\citeauthoryear{ZuHone et al.}{2016}]{ZuHone2016}
ZuHone, J. A., Markevitch, M., \& Zhuravleva, I., 2016, ApJ, 817, 110








\end{thebibliography}
\end{document}